\documentclass[peerreview]{IEEEtran}
\usepackage{cite} 
\usepackage{url} 
\usepackage[utf8]{inputenc} 
\usepackage{booktabs} 
\usepackage{graphicx}
\usepackage{longtable,tabularx}
\usepackage{listings}
\usepackage{xcolor}
\usepackage{float}
\usepackage{color, colortbl}
\definecolor{Gray}{gray}{0.8}

\begin{document}
\begin{titlepage}
    \centering
    \vspace*{2cm}
    
    {\Huge \textbf{Implementation of the Collision Avoidance System for DO-178C Compliance} \par}
    \vspace{1.5cm}
    
    {\Large Technical Report Number: POLY\_HESL\_RT\_19092025 \par}
    \vspace{1cm}
    
    {\large
    Rim Zrelli\textsuperscript{1,*}, 
    Henrique Amaral Misson\textsuperscript{1,*}, \\
    Sorelle Kamkuimo\textsuperscript{2}, 
    Maroua Ben Attia\textsuperscript{3}, \\
    Abdo Shabah\textsuperscript{3}, 
    Felipe Gohring de Magalhaes\textsuperscript{1}, 
    Gabriela Nicolescu\textsuperscript{1} \par}
    
    \vspace{1cm}
    {\small
    \textsuperscript{1} Department of Computer Engineering and Software Engineering, Polytechnique Montréal \\
    \textsuperscript{2} University of Quebec in Chicoutimi \\
    \textsuperscript{3} HumanITas Solution \\
    }
    
    \vfill
    {\large Date: September 20, 2025 \par}
    
    \vspace{1cm}
    \rule{0.6\linewidth}{0.5pt}\\
    \small{* Both authors contributed equally to this research. \\
    \textsuperscript{2} The author is now affiliated with the School of Digital Arts, Animation and Design at the University of Québec at Chicoutimi.}
\end{titlepage}

\clearpage
\thispagestyle{empty}
\vspace*{\fill}
\begin{center}
\small
\textbf{Suggested citation - Technical report}\\[0.75em]
R. Zrelli, H. A. Misson, S. Kamkuimo, M. B. Attia, A. Shabah, F. G. de Magalhães and G. Nicolescu, \textit{" Implementation of the Collision Avoidance System for DO-178C Compliance"}, 
Technical Report \textsc{POLY\_HESL\_RT\_19092025}, Polytechnique Montréal, 2025. [En ligne]
\end{center}
\vspace*{\fill}
\clearpage

\title{Implementation of the Collision Avoidance System for DO-178C Compliance}


\author{
    \IEEEauthorblockN{
        Rim Zrelli\textsuperscript{1,*},
        Henrique Amaral Misson\textsuperscript{1,*},
        Sorelle Kamkuimo\textsuperscript{2},
        Maroua Ben Attia\textsuperscript{3},
        Abdo Shabah\textsuperscript{3},
        Felipe Gohring de Magalhaes\textsuperscript{1},
        Gabriela Nicolescu\textsuperscript{1}
    }
    
    \IEEEauthorblockA{\textsuperscript{1} Department of Computer Engineering and Software Engineering, Polytechnique Montréal, QC, Canada} \\
    \IEEEauthorblockA{\textsuperscript{2} University of Quebec in Chicoutimi, Saguenay, QC, Canada} \\
    \IEEEauthorblockA{\textsuperscript{3} HumanITas Solutions, Montréal, QC, Canada} \\
    \textsuperscript{*}Equal contribution
}
\date{25/4/15}

\maketitle
\tableofcontents
\listoffigures
\listoftables

\IEEEpeerreviewmaketitle
\begin{abstract}
This technical report presents the detailed implementation of a Collision Avoidance System (CAS) for Unmanned Aerial Vehicles (UAVs), developed as a case study to demonstrate a rigorous methodology for achieving DO-178C compliance in safety-critical software. The CAS is based on functional requirements inspired by NASA's Access 5 project and is designed to autonomously detect, evaluate, and avoid potential collision threats in real-time, supporting the safe integration of UAVs into civil airspace.

The implementation environment combines formal methods, model-based development, and automated verification tools, including Alloy, SPIN, Simulink Embedded Coder, and the LDRA tool suite. The report documents each phase of the software lifecycle: requirements specification and validation, architectural and detailed design, coding, verification, and traceability, with a strong focus on compliance with DO-178C Design Assurance Level B objectives.

Results demonstrate that formal modelling and automated toolchains enabled early detection and correction of specification defects, robust traceability, and strong evidence of verification and validation across all development stages. Static and dynamic analyses confirmed code quality and coverage, while formal verification methods provided mathematical assurance of correctness for critical components. Although the integration phase was not fully implemented, the approach proved effective in addressing certification challenges for UAV safety-critical systems.

\end{abstract}

\section{Introduction}
This technical report documents the implementation of a Collision Avoidance System (CAS) for Unmanned Aerial Vehicles (UAVs), developed as a case study to demonstrate the application of a methodology aimed at achieving DO-178C compliance for safety-critical software. The underlying methodology, including its formal foundations, modelling strategy, and toolchain integration, is the subject of a separate journal article.

The purpose of this report is to provide a complete and detailed account of the development and verification activities carried out throughout the software lifecycle of the CAS, following the objectives and practices defined in the DO-178C standard. It includes all specification, modelling, coding, verification, and traceability steps performed using formal methods and automated tools.

This report is intended as a complement to the methodology paper, offering supporting implementation evidence and serving as a practical reference for readers seeking a concrete example of how the proposed approach can be applied to a real-world, safety-critical UAV software system.

\section{System Overview}
The CAS for UAVs is a critical technology designed to ensure the safe and reliable integration of autonomous UAVs into the airspace by autonomously detecting, evaluating, and avoiding potential collision threats in real-time. As UAV usage increases across commercial, civil, and defence sectors, safely managing airspace congestion becomes paramount. CAS addresses these challenges by autonomously replicating and enhancing the traditional "see-and-avoid" capabilities typically performed by human pilots, enabling UAVs to operate securely within dynamic airspace environments.

Inspired by NASA’s Access 5 Project, specifically the "Collision Avoidance Functional Requirements for Step 1" document \cite{NASA_CA_Reqs}, the CAS implementation leverages proven methodologies for cooperative collision avoidance originally developed for high altitude, long endurance unmanned aircraft systems. NASA's defined functional requirements include detection of traffic, tracking, collision potential evaluation, threat prioritization, determination of avoidance maneuvers, maneuver command, and execution.

The CAS continuously monitors UAV surroundings, detecting cooperative aircraft and objects within defined surveillance volumes. Its importance is especially pronounced during Beyond Visual Line of Sight (BVLOS) operations, where human operators lack direct visual contact. CAS autonomously assesses collision risks, prioritizes identified threats, and determines appropriate avoidance maneuvers, ensuring UAVs maintain safe flight trajectories. This autonomous decision-making capability is crucial when operating in shared airspaces alongside manned aircraft, requiring accurate, timely, and data-driven collision avoidance decisions.

Under the guidelines provided by the NASA Access Project, CAS is designed to achieve a robust performance in cooperative environments. According to the NASA Document, cooperative collision avoidance functionality encompasses the following core tasks:

\begin{itemize}
    \item Traffic Detection: The system continuously monitors the UAV's surroundings, detecting both cooperative and non-cooperative aircraft or objects.
    \item Track Traffic: The previous traffic detected is tracked to establish a reliable track history for each object.
    \item Collision Risk Assessment: Once traffic is detected, the CAS evaluates the potential for a collision based on trajectory predictions, considering speed, direction, and proximity.
    \item Threat Prioritization: In situations where multiple collision threats are identified, the system prioritizes these threats based on proximity and the immediacy of potential impact.
    \item Avoidance Maneuver Determination: CAS autonomously determines the most appropriate avoidance maneuver (e.g., altitude adjustment or lateral deviation) to prevent a collision with the highest-priority threat.
    \item Maneuver Command: The system sends maneuver commands to the UAV’s flight control system to execute the avoidance action in real-time.
    \item Maneuver Execution: Executes commanded maneuvers immediately through the UAV’s flight control systems.
\end{itemize}

Compliance considerations emphasize the fundamental role of systematic safety evaluations as defined in standards like ARP4761. These evaluations are critical to ensuring the Collision Avoidance System adheres to the rigorous safety and reliability requirements outlined by the DO-178C standard. As a Design Assurance Level B (DAL-B) system, CAS demands stringent software development, verification, and validation processes to reliably mitigate hazardous scenarios, thereby preventing severe consequences such as mid-air collisions or ground impacts.

Through alignment with NASA’s established guidelines and rigorous adherence to DO-178C, CAS significantly enhances UAV operational safety and compliance, facilitating their seamless integration into increasingly crowded airspaces.

\section{Implementation Environment}
The CAS implementation detailed in this report was carried out within a structured software development environment designed explicitly to meet DO-178C certification requirements at the DAL-B. The development process employed an iterative V-model lifecycle approach, systematically covering the phases of requirements, design, coding, and integration.

The implementation leveraged a carefully selected combination of formal methods, modelling tools, and automated verification suites to ensure thorough compliance with DO-178C standards:

\begin{itemize}
    \item \textbf{Alloy Analyzer}: Utilized for formal specification and rigorous verification of system requirements. Alloy was essential for modelling SRATS (System Requirements Allocated to Software), High-Level Requirements (HLRs), and Low-Level Requirements (LLRs). It allowed for the early detection of ambiguities, inconsistencies, and errors through formal specification.

    \item \textbf{SPIN Model Checker (Promela)}: SPIN was employed for formal verification of design models. It provided exhaustive checking of system behaviours and facilitated the validation of critical properties and interactions defined during the software design phase.

    \item \textbf{LDRA Tool Suite}
        \begin{itemize}
            \item \textbf{TBmanager}: Managed requirements throughout the lifecycle, ensuring comprehensive traceability from initial SRATS through HLRs, LLRs, to code and tests. TBmanager automated the generation of traceability matrices and facilitated consistent documentation for certification purposes.

            \item \textbf{TBvision}: Performed static code analysis, verifying adherence to coding standards and identifying potential issues early in the coding phase. It also provided structural coverage analysis for statement and decision.

            \item \textbf{TBrun}: Supported dynamic testing, automatically generating and executing test cases based on detailed requirements. TBrun ensured unit-level and integration-level testing were executed comprehensively, and all test results were documented and traceable back to the respective requirements.
            
        \end{itemize}

    \item \textbf{Simulink Embedded Coder}: Used for automatic code generation from validated Simulink design models. This automated process minimized manual coding errors and ensured alignment between the software architecture and executable code, directly supporting DO-178C compliance objectives.

    \item \textbf{Architecture Analysis \& Design Language (AADL)}: Provided formal modelling of the CAS software architecture. AADL facilitated clear documentation of system interfaces, data flows, and interactions among subsystems such as Traffic Detection, Tracking, Collision Evaluation, Threat Prioritization, Maneuver Determination, and Maneuver Command.
\end{itemize}

\section{Requirements Phase Implementation}
The first phase of the methodology focuses on defining and evaluating the CAS requirements for DO-178C compliance. This entailed identifying the SRATS, including functional and non-functional aspects critical to CAS performance and safety. To ensure these requirements were clear, unambiguous, and precise, a formal model of the SRATS was developed using the Alloy language, and the Alloy Analyzer was used to conduct rigorous consistency checks.

A notable example is SRATS\_002, which specifies the spatial boundaries of CAS surveillance based on detection range, azimuth, and elevation fields of regard. The Alloy model for SRATS\_002 is as follows:
\lstdefinestyle{Alloy}{
    basicstyle=\ttfamily\small,
    keywordstyle=\color{blue}\bfseries,
    commentstyle=\color{gray}\itshape,
    stringstyle=\color{red},
    numbers=left,
    numberstyle=\tiny\color{gray},
    stepnumber=1,
    numbersep=10pt,
    breaklines=true,
    frame=lines,
    captionpos=b,
    showstringspaces=false
}

\begin{lstlisting}[style=Alloy, caption={Alloy Specification for SRATS\_002}, label={lst:AlloySrats2}]
    fact SRATS_002 {
    all sys: CollisionAvoidanceSystem | 
        sys.surveillanceVolume.traffic = { t: Traffic |
            t.x <= sys.detectionRange and
            (t.y >= sys.minAzimuthFieldOfRegard and t.y <= sys.azimuthFieldOfRegard) and
            (t.z >= sys.minElevationFieldOfRegard and t.z <= sys.elevationFieldOfRegard)
        }
}
\end{lstlisting}
This Alloy model formally encapsulates the spatial constraints of the CAS surveillance volume. Assertions, such as \textit{CompleteTrafficDetection}, were used to verify that all traffic meeting these criteria was correctly detected. The Alloy Analyzer validated this assertion without counterexample, confirming the logical consistency of the system model under the tested scenarios.

During the verification process of other assertions, Alloy Analyzer discovered several inconsistencies in the SRATS, revealing issues such as incorrect assumptions about traffic detection range and ambiguities in threat prioritization. These findings were documented as defects and reported to the system engineering team for correction. Each issue was addressed using an iterative refinement strategy, resulting in a more accurate and reliable set of system requirements consistent with the desired system behaviour and safety objectives. A comprehensive list of SRATS is provided in Appendix \ref{App:srats} for reference.

Following the refinement and verification of the SRATS, the validated requirements were imported into LDRA TBManager to support the HLR development. The SRATS were systematically decomposed into key CAS functions, enabling the creation of software-level HLRs (detailed in Appendix \ref{App:hlr}) that directly traced back to the system requirements.

\begin{figure}[hbt!]
\centering
\includegraphics[width=0.5\textwidth]{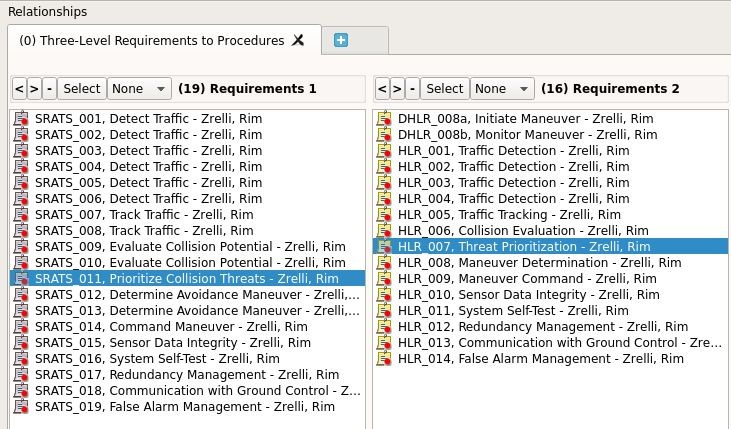}
\caption{TBmanager relationship view between SRATS and HLR}
\label{fig:HLR_TBManager}
\end{figure}

The initial draft of the HLRs was developed and set up within a versioned baseline in the tool. As depicted in Figure \ref{fig:HLR_TBManager}, a traceability was created indicating that each HLR was directly linked to its associated SRATS. Following the outlined development activities, the next stage was to ensure that the requirements fulfilled DO-178C objectives, specifically those listed in \textit{Table A-3 ("Verification of Outputs of Software Requirement Process")} of the DO-178C.

To ensure that the HLRs adhered to DO-178C objectives for compliance with SRATS, accuracy, consistency, and verifiability, a formal model was constructed in Alloy. This model enabled the detection of inconsistencies and ambiguities, facilitating early refinement and validation of the requirements.

\begin{figure}[hbt!]
\centering
\includegraphics[width=0.49\textwidth]{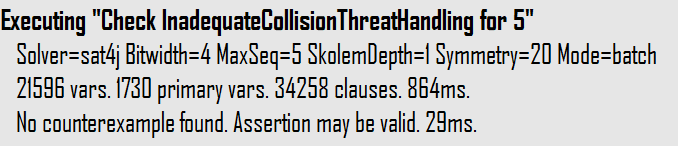}
\caption{Results for checking the InadequateCollisionThreatHandling assertion after refinement}
\label{fig:HLRVerif}
\end{figure}

The verification process involved:
\begin{itemize}
    \item Initial Model Execution: The initial Alloy execution demonstrated that the HLRs were logically consistent and free of major contradictions. However, one critical assertion, \textit{InadequateCollisionThreatHandling}, failed, exposing gaps in HLR\_008 and HLR\_009, which relate to the handling of collision threats and associated maneuvers.
    
    \item Analysis of the counterexample: The failed assertion revealed a scenario where identified collision threats did not trigger corresponding avoidance maneuvers. This gap was traced to ambiguities in the activation and prioritization conditions for maneuvers:
    \begin{itemize}
        \item Ambiguity in maneuver activation: HLR\_008 lacked specificity about when maneuvers should be initiated for detected threats.
        \item Gaps in prioritization logic: HLR\_009 did not define how to prioritize maneuvers when multiple threats were present.
        \item Dynamic adjustment issues: Conditions for dynamically updating or terminating maneuvers were insufficiently detailed.
    \end{itemize}
    
    \item Refinement: Based on this analysis, HLR\_008 and HLR\_009 were refined
    to explicitly mandate the initiation of maneuvers for all detected threats unless overridden by a higher-priority threat and to address maneuver termination and prioritization. To bridge the gap between HLRs and LLRs, two new DHLRs (DHLR\_008a and DHLR\_008b) and their rationale were introduced to clarify the conditions for handling collision threats. 
    
    \item Reverification: The refined HLRs and the DHLRs were re-modelled in Alloy. Assertions, including \textit{InadequateCollisionThreatHandling}, were verified (as depicted in Figure \ref{fig:HLRVerif}), and no further counterexamples were found. This confirmed compliance with DO-178C objectives A3.2 (accuracy), A3.4 (verifiability), and A3.5 (conformity to standards).
\end{itemize}

The assertion \textit{InadequateCollisionThreatHandling} states that some traffic detected as a threat is not being actively maneuvered against when it should be and it is specified as follows:

\begin{lstlisting}[style=Alloy, caption={InadequateCollisionThreatHandling Assertion}, label={lst:AseertionHLR}]
assert InadequateCollisionThreatHandling{
    all sys: CollisionAvoidanceSystem, t: Traffic | 
        t in sys.collisionThreats implies (
            some m: sys.maneuvers | 
            m.isActive = True and 
            t in m.threat.spaceZone and 
            (no t2: sys.collisionThreats - t | t2.threatLevel > t.threatLevel and t2.timeToCollision < t.timeToCollision)
        )
}
check InadequateCollisionThreatHandling for 5
\end{lstlisting}

\begin{figure}[hbt!]
\centering
\includegraphics[width=0.5\textwidth]{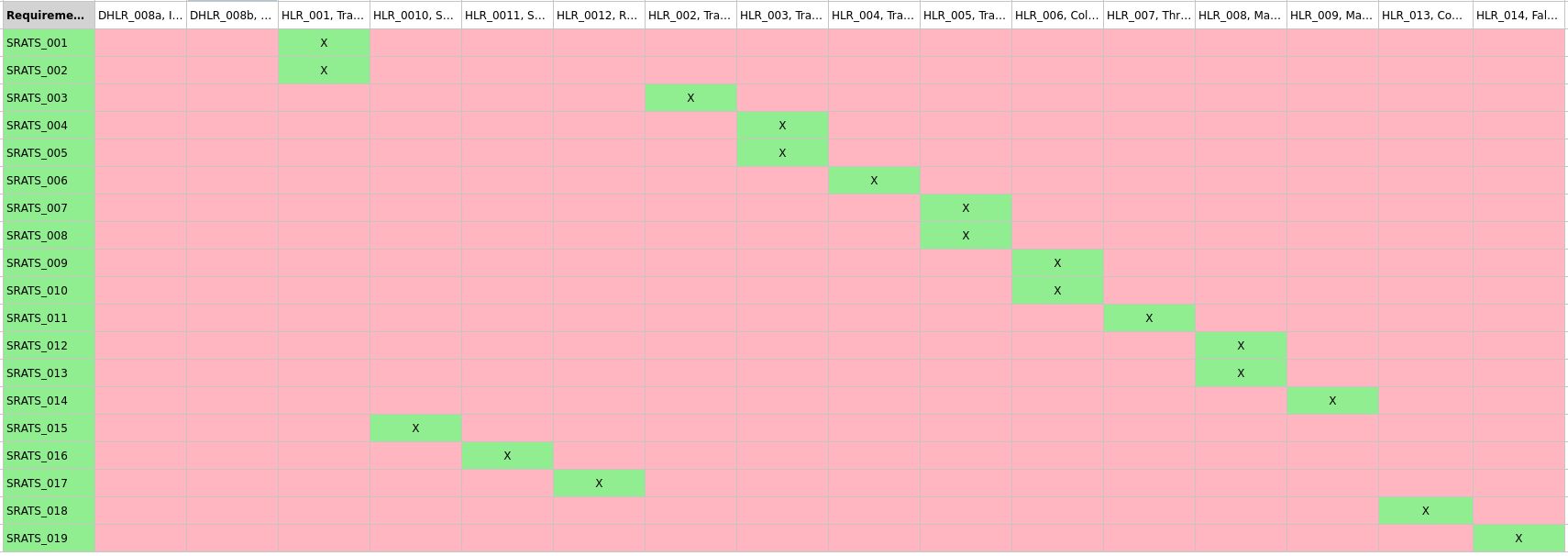}
\caption{Traceability matrix between SRATS and HLRs}
\label{fig:HLR_MatrixTrace}
\end{figure}

The validated HLRs were imported back into TBManager to finalize the traceability matrix and link each HLR and DHLR to its corresponding SRATS. This integration facilitated the generation of certification artifacts, such as the Software Requirements Traceability Matrix (SRTM) (Figure \ref{fig:HLR_MatrixTrace}).

The final activity of the Requirements phase, referred to as \textit{Requirements Specification and Validation}, concentrates on validating the HLRs to ensure their robustness and readiness for further development phases. Previously, a set of HLRs was drafted using the SRATS, and these HLRs underwent preliminary verification activities. In this phase, however, a peer review is carried out by an independent reviewer who was not involved in the original HLR drafting. This impartial evaluation is crucial to ensuring that all standards have been satisfied, that the requirements are accurate and unambiguous and that there is complete traceability between each HLR and the system requirements, which includes all relevant parts.

If any issues or inconsistencies are discovered during the peer review, the original author logs a defect for rectification, which initiates an iterative process to enhance the requirements. The reviewer follows a rigorous checklist to confirm that each DO-178C objective is met, including the standards' clarity, consistency, and completeness. This systematic approach ensures that the HLRs meet stringent quality and compliance criteria, reducing the likelihood of errors that could spread to later phases of development. Once all of the checklist elements have been satisfactorily completed, the evaluation is formally documented.

\begin{figure}[htb]
\centering
\includegraphics[width=0.5\textwidth]{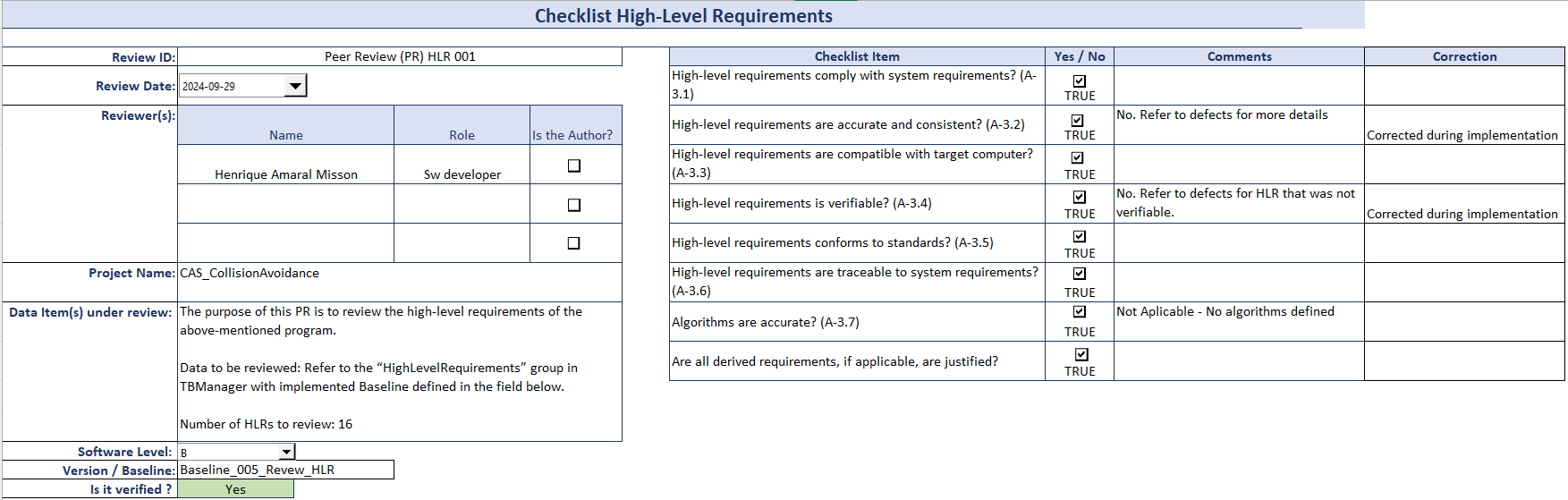}
\caption{Checklist used for peer review of HLRs during the validation}
\label{fig:HLR_PeerReview}
\end{figure}

Figure \ref{fig:HLR_PeerReview} illustrates the peer review checklist used during this validation phase. This checklist ensures that all DO-178C objectives related to HLRs are rigorously evaluated, including compliance with system requirements (A3.1), accuracy and consistency (A3.2), and traceability (A3.6). Any identified issues are documented in the "Comments" section and addressed iteratively during the correction process.

Throughout this process, configuration management is strictly maintained in TBManager, where all review activities, changes, baselines, and defects are recorded. This allows for full documentation of the requirements development process, and TBManager can provide a report summarizing the results of the \textit{Requirements Specification and Validation} phase. The final output of this phase, known as Requirements Data, consists of both validated HLRs and DHLRs, which will serve as the foundational specifications for the design and coding phases.

\begin{figure}[htb]
\centering
\includegraphics[width=0.5\textwidth]{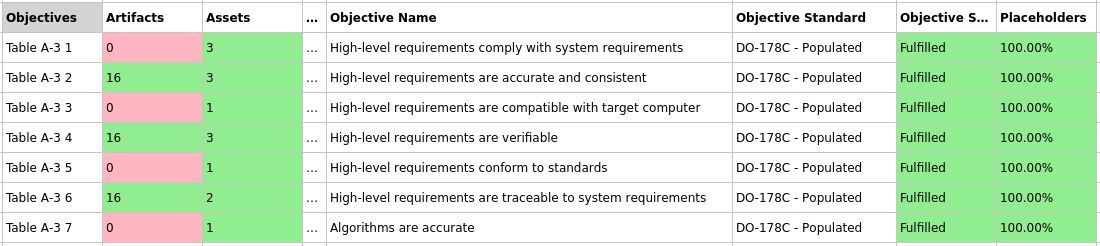}
\caption{DO-178C objectives for verification of outputs of software requirement process in TBmanager}
\label{fig:objectiveA3}
\end{figure}

The requirements phase concluded with the generation of validated Requirements Data, consisting of SRATS, HLRs, and DHLRs. These outputs formed a robust foundation for the design and coding phases, fulfilling all Table A-3 objectives of DO-178C (Figure \ref{fig:objectiveA3}). This iterative and tool-supported process mitigated risks early, ensuring that subsequent phases adhered to the highest standards of safety and certification rigour.

\section{Design Phase Implementation}

The design phase of the methodology centred around defining the architecture and the LLRs of the CAS to ensure that it adhered to both operational and safety goals, as outlined in the HLRs. 

The design phase begins with an in-depth review of the \textit{Requirements Data} generated during the requirements phase. The goal of this analysis is to decompose the HLRs into modular subsystems, each responsible for a critical function of the CAS. Key subsystems identified include traffic detection, tracking, collision evaluation, threat prioritization, and maneuver determination. For each subsystem, essential inputs, outputs, and interactions with other components and external equipment are carefully described and documented. Once these components and their interfaces have been identified, a preliminary architecture model is developed. This model, created in the Architecture Analysis and Design Language (AADL) \cite{feiler2004overview} using the Osate tool \cite{feiler2004open}, encapsulates both internal relationships and software-level views of CAS components.

\begin{figure*}[hbt!]
\centering
\includegraphics[width=1\textwidth]{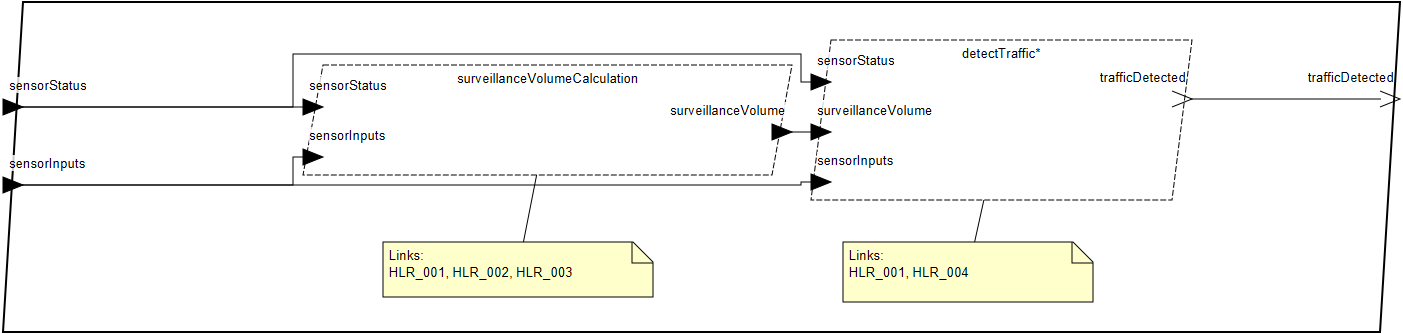}
\caption{AADL model of the traffic detection subsystem}
\label{fig:TrafficDetectionAADL}
\end{figure*}

One of the critical subsystems identified during the design phase is \textit{Traffic Detection}. This subsystem is responsible for continuously monitoring the UAV's surroundings. The architecture of this subsystem was modelled in AADL, as shown in Figure \ref{fig:TrafficDetectionAADL}, capturing its inputs (sensor data from onboard radar and external data sources), outputs (a list of detected traffic), and interactions with other CAS components, such as \textit{Traffic Tracking} and \textit{Collision Evaluation}.

The architectural representation (Figure \ref{fig:TrafficDetectionAADL}) encapsulates the core operations of this subsystem, including surveillance volume calculation and traffic detection. The figure highlights traceability to the associated HLRs (HLR\_001 - HLR\_004), ensuring that the subsystem aligns with the requirements established in the previous phase. For a detailed view of the architectural diagrams of all the identified components of the CAS Architecture refer to Appendix \ref{App:llr}.

\begin{figure}[hbt!]
\centering
\includegraphics[width=0.5\textwidth]{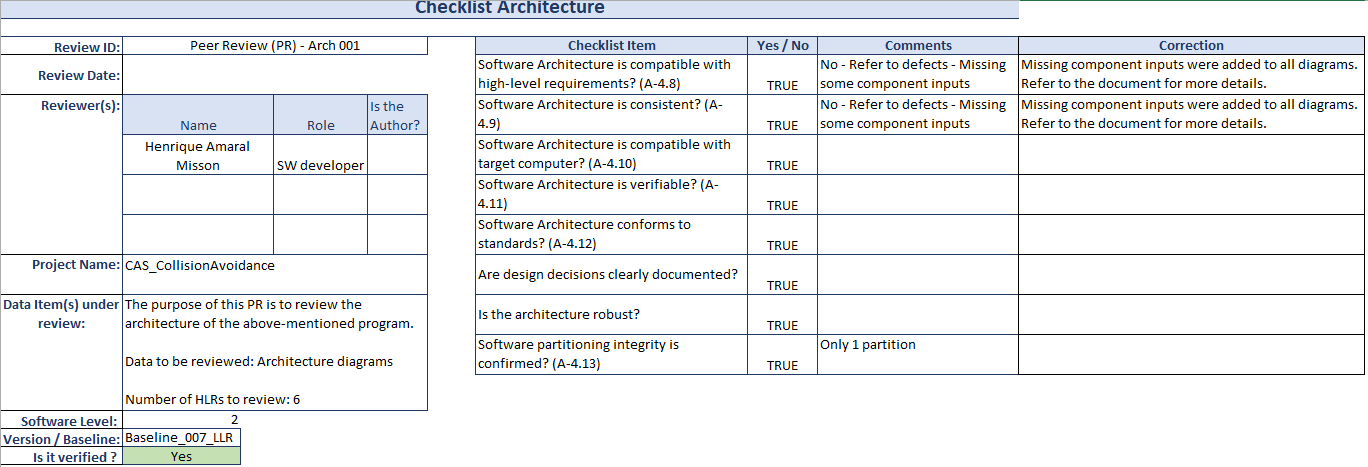}
\caption{Checklist used for peer review of the architecture}
\label{fig:Archi_PeerReview}
\end{figure}

An independent peer review was conducted to confirm that the architecture aligns with the HLRs and meets the DO-178C objectives. This review involved a detailed checklist evaluation to verify the architecture's compliance with key DO-178C objectives, including consistency, verifiability, and traceability to HLRs. Figure \ref{fig:Archi_PeerReview} illustrates the peer review checklist completed during this validation phase. As shown, any defects, such as missing component inputs, were identified and corrected during this process. 

In addition, formal verification was performed using model-checking approaches to ensure the architecture's accuracy and reliability. The architecture was modelled in the Promela language and analyzed using the Spin model checker, which converts features of interest into LTL statements. These LTL attributes specify important system behaviours and limitations, guaranteeing that the architecture design meets the necessary safety and functional requirements. The verification process focused on critical system properties, such as liveness properties, ensuring that critical actions (such as avoidance maneuvers) are eventually executed when necessary and safety properties, preventing the activation of incorrect or dangerous maneuvers, thereby avoiding unintended consequences.

Key properties analyzed through LTL claims included:
\begin{itemize}
    \item c1: A maneuver command is eventually executed for detected collision threats. \\ $[]<>CommandManeuver@SendCmdTrue$.
    \item c2: Detection of traffic leads to a maneuver command. \\ $[](DetectTraffic@SendTrafficDetected) \rightarrow <> (CommandManeuver@SendCmdTrue)$.
    \item c3: Collision evaluation follows traffic detection. \\ $[](DetectTraffic@SendTrafficDetected)\rightarrow <>(CollisionEvaluation@EvaluateCollisionPotential)$.
\end{itemize}

The results showed that no assertion violations or invalid end states were detected, demonstrating that the model satisfies all safety and functional properties in the explored scenarios. However, some claims exhibited unreached states. The analysis of these results indicates that while the core architecture is robust, the unreached states suggest potential areas for refinement or the need for additional test scenarios. For example, claim \textit{c1} revealed missed conditions where detected threats did not lead to a maneuver command, suggesting the need for enhanced conditions in the model.
Claim \textit{c3} highlighted edge cases where collision evaluation did not consistently follow traffic detection, pointing to potential gaps in the prioritization logic. These findings were documented and addressed iteratively, ensuring continuous improvement of the architecture model and alignment with the safety-critical objectives of DO-178C.

While architectural design activities occur outside the LDRA tool suite, the verified architecture model and supporting documentation were imported into TBManager. This integration ensured that all verification results were tracked and included in the overall configuration management system, facilitating traceability and certification artifact generation.

As part of the iterative refinement process, the LLR development activities focused on translating the HLRs into detailed, comprehensive specifications that would guide the coding phase. These LLRs were precisely linked to planned software implementations, enabling developers to write functions with exact behaviour and data flow specifications. Each LLR ensured traceability to the HLRs and architecture while meeting DO-178C compliance objectives.

The LLRs were derived to define the operational behaviour of each subsystem with precision. For instance:
\begin{itemize}
    \item LLR-001 specifies that when the sensor status is active, the TrafficDetection function retrieves the sensor inputs, initiating the data acquisition process.
    \item LLR-002 establishes the rules for validating sensor inputs, ensuring parameters such as DetectionRange, AzimuthFOR, and ElevationFOR fall within safe operational ranges.
\end{itemize}

A comprehensive list of the LLRs is provided in Appendix \ref{App:llr}, which includes LLR-001 through LLR-032.

\begin{figure}[hbt!]
\centering
\includegraphics[width=0.5\textwidth]{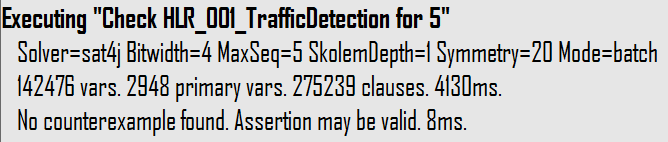}
\caption{Results for checking the HLR\_001\_TrafficDetection assertion}
\label{fig:LLRVerif}
\end{figure}

Formal verification was conducted to validate the accuracy, consistency, and compliance of the LLRs with the HLRs. This process, conducted in two stages, ensured adherence to DO-178C objectives A4.1 (compliance), A4.2 (accuracy), and A4.4 (verifiability).
\begin{enumerate}
    \item Execution of the Alloy model representing all LLRs: \\
    A comprehensive Alloy model was created to represent all LLRs as a cohesive system. This model encoded the functional logic, interactions, and constraints described in the LLRs. Using the Alloy Analyzer, the model was executed to validate internal consistency and interaction across subsystems. The analysis confirmed that the LLRs collectively formed a logically sound design. Instances were generated successfully without contradictions, demonstrating that the specified constraints did not conflict and that subsystems interacted as expected under the architectural model.
    \item Assertion-based compliance checks with HLRs: \\
    Once the internal consistency of the LLRs was validated, assertions were defined for each HLR to ensure compliance. These assertions encapsulated key behaviours derived from the HLRs and were checked against the Alloy model. For example:
    \begin{itemize}
        \item Assertion HLR\_001\_TrafficDetection verified that the inputs conformed to the correct \textit{FOR} (Field of Regard), such as Azimuth FOR and Elevation FOR, for detecting traffic. It is formally expressed as:
\begin{lstlisting}[style=Alloy, caption={HLR\_001\_TrafficDetection Assertion}, label={lst:AseertionLLR}]
assert HLR_001_TrafficDetection {
    all td: TrafficDetection |
        some td.sensorInput =>
        td.sensorInput.DetectionRange > 0 and 
        td.sensorInput.AzimuthFOR >= -110 and td.sensorInput.AzimuthFOR <= 110 and 
        td.sensorInput.ElevationFOR >= -15 and td.sensorInput.ElevationFOR <= 15
    }
    check HLR_001_TrafficDetection for 5
\end{lstlisting}
        The Alloy Analyzer executed this assertion, and no counterexamples were found (as shown in Figure \ref{fig:LLRVerif}), confirming compliance with HLR\_001. The absence of counterexamples demonstrated that the system adhered to operational parameters for the sensor input's field of regard, ensuring comprehensive spatial coverage and reliable traffic detection.
        
        \item Assertion HLR\_008\_ManeuverDetermination ensured that appropriate maneuvers were determined in response to detected collision threats. The validation of this assertion highlighted the system's ability to proactively determine maneuvers upon detecting potential collisions.
    \end{itemize}
    The results of these checks confirmed that all assertions were validated successfully, with no counterexamples identified. The LLRs adhered to their corresponding HLRs, ensuring alignment with operational and safety goals. These results validated the robustness of the LLRs and their compliance with DO-178C objectives, minimizing risks of errors propagating into the coding phase.
\end{enumerate}

Throughout the design phase, automated traceability tools (TBmanager) were employed to ensure that all LLRs could be traced back to their respective HLRs. This traceability guaranteed that any changes made during the design process were reflected throughout the system, maintaining alignment with DO-178C objective A4.6 (traceability). TBmanager was used to generate and maintain traceability matrices, linking HLRs to their corresponding LLRs. These matrices ensured that each requirement was traceable from its specification to its detailed design. As refinements were made during verification activities, the matrices were continuously updated to reflect the most current relationships between HLRs and LLRs. The integration of automated tools allowed for real-time compliance checks against DO-178C requirements. This approach ensured that the evolving design consistently adhered to certification standards, minimizing the risk of misalignment between HLRs and LLRs during iterative updates.

\begin{figure}[hbt!]
\centering
\includegraphics[width=0.5\textwidth]{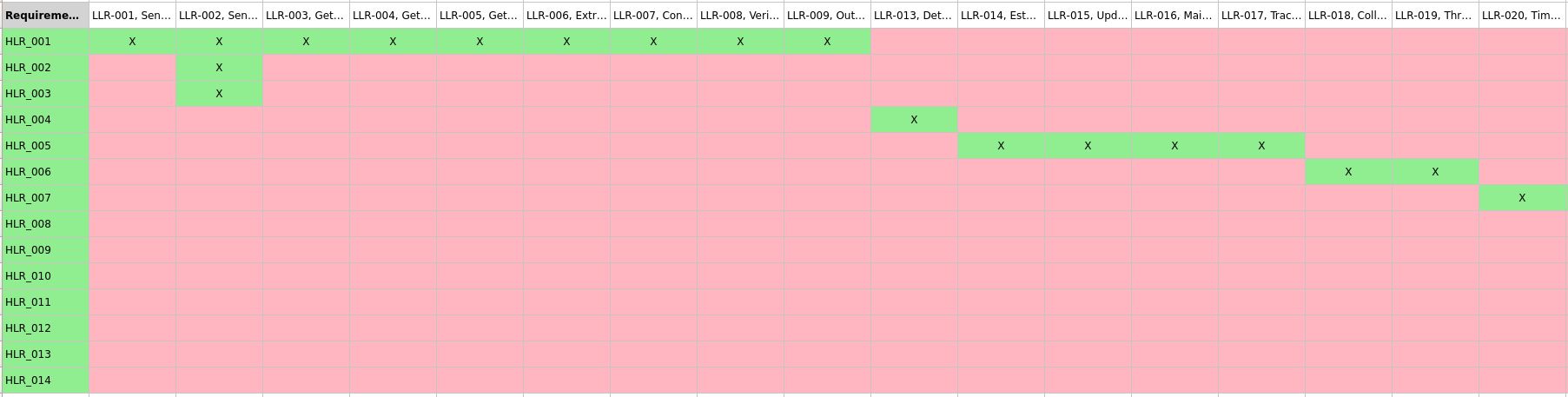}
\caption{Traceability matrix between HLRs and LLRs}
\label{fig:LLR_MatrixTrace}
\end{figure}

Figure \ref{fig:LLR_MatrixTrace} illustrates the traceability matrix generated, which links HLRs to their associated LLRs. The green markers indicate relationships between requirements, ensuring complete coverage and traceability as mandated by DO-178C standards.

The FMEA and FTA activities were crucial in identifying potential failure points and assessing their impact on the CAS. These analyses strengthened the robustness and reliability of the system, uncovering failure modes that could compromise safety and functionality.

The FMEA focused on identifying potential failure modes across each subsystem of the CAS, evaluating their causes, and analyzing their effects. This systematic approach prioritized failure modes based on their severity and proposed mitigations to address vulnerabilities. For example, in the \textit{Traffic Detection subsystem}, failure modes such as \textit{“Traffic detected late”} or \textit{“Traffic detected at incorrect height”} were analyzed. These failures were assessed for their impact on the UAV's ability to respond to collision threats in time. Recommendations included implementing dual-sensing technologies, periodic sensor calibration, and improving data validation protocols (see Appendix \ref{App:fmea} for the complete FMEA report).

\begin{figure}[hbt!]
\centering
\includegraphics[width=0.5\textwidth]{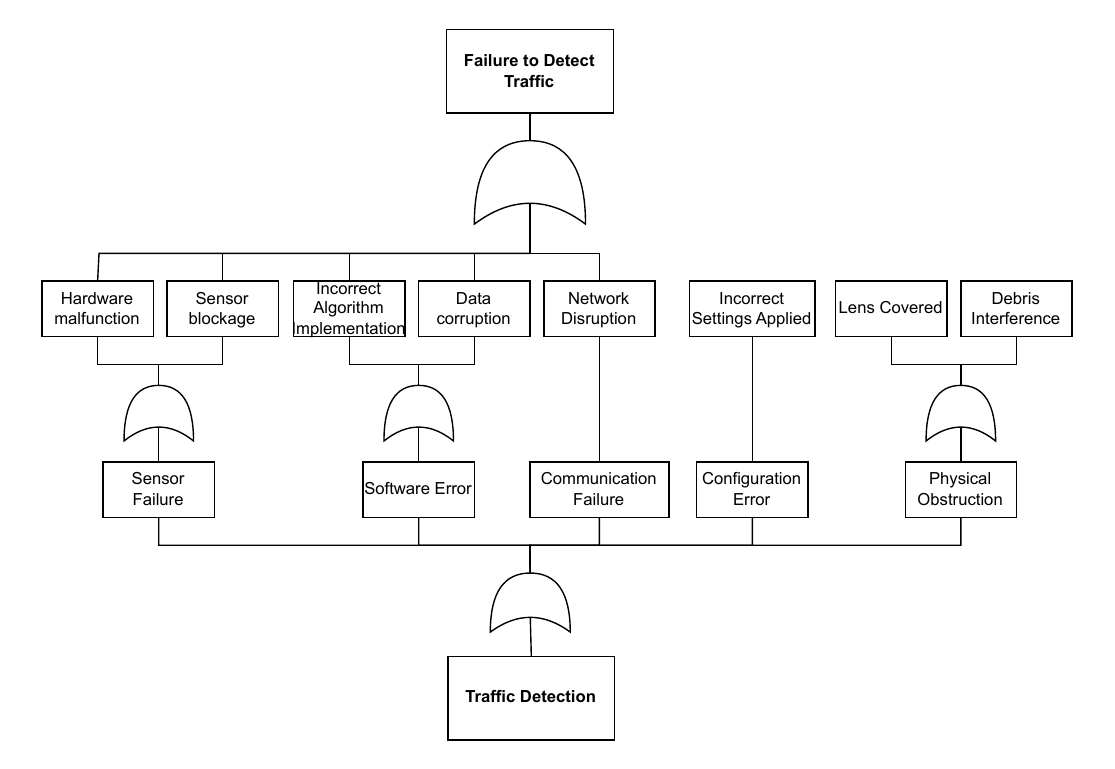}
\caption{FTA for the Traffic Detection subsystem}
\label{fig:FTA_TrafficDetection}
\end{figure}

Complementing the FMEA, the FTA provided a graphical representation of fault scenarios and their contributing factors. By tracing failures back to their root causes, the FTA highlighted dependencies and vulnerabilities in the system architecture. For the \textit{Traffic Detection} subsystem, the FTA (Figure \ref{fig:FTA_TrafficDetection}) identified key contributors to detection failures, including sensor failure, software errors, communication failures, configuration errors, and physical obstruction.

The final activity in the design phase involved a comprehensive review of the CAS architecture and design documentation to ensure conformance to all safety and quality standards. This step aimed to validate the readiness of the system design for the subsequent coding phase, ensuring that all critical safety and operational requirements were met.

To verify the completeness, accuracy, and consistency of the design documentation, independent peer reviews were conducted. These reviews focused on validating that:
\begin{itemize}
    \item The design aligned with the safety-critical objectives outlined in DO-178C.
    \item All design artifacts were traceable to their respective HLRs and LLRs.
    \item The architecture and LLRs were robust and sufficiently detailed to guide the coding phase.
\end{itemize}
Like the HLRs and architecture reviews, the peer reviews employed detailed checklists to ensure systematic evaluation.

\begin{figure}[hbt!]
\centering
\includegraphics[width=0.5\textwidth]{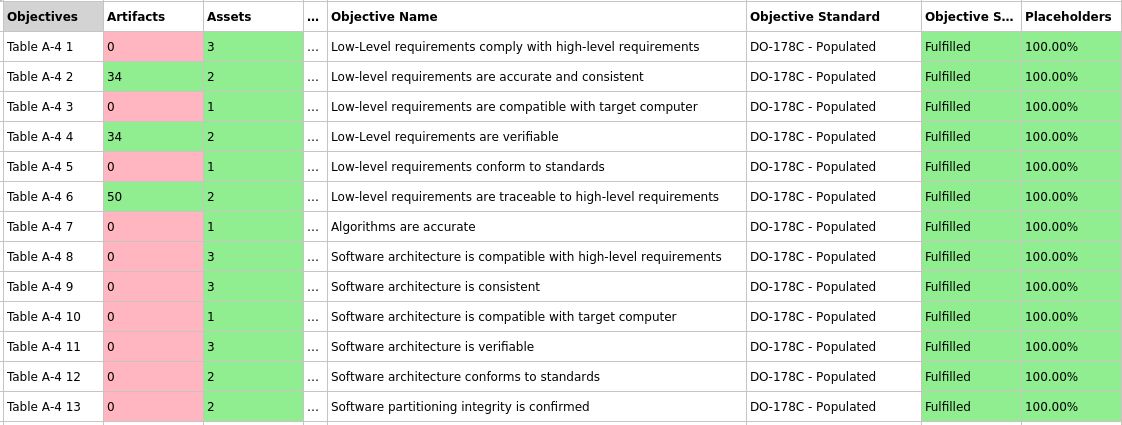}
\caption{DO-178C objectives for verification of outputs of the software design process in TBmanager}
\label{fig:ObjectivesA4}
\end{figure}

By the conclusion of this phase, all defects identified during the reviews and verification activities were resolved. Figure \ref{fig:ObjectivesA4} presents a summary of the DO-178C objectives fulfilled during the design phase. This table, generated using TBManager, demonstrates 100\% compliance with objectives related to LLRs and software architecture, including traceability, consistency, and verifiability. The robust alignment between design activities and certification requirements further validated the readiness of the design for the coding phase.

By the end of the design phase, the CAS architecture and LLRs have been thoroughly verified and validated, ensuring that they meet all necessary safety and operational requirements. This provided a robust foundation for the subsequent coding phase, where the verified design would be translated into executable code.

\section{Coding Phase Implementation}
The Coding phase of the software development process focuses on translating the system design into executable source code. Given the critical nature of the software, a hybrid method was used, combining automated code creation with manual development. 

\begin{figure}[hbt!]
\centering
\includegraphics[width=0.5\textwidth]{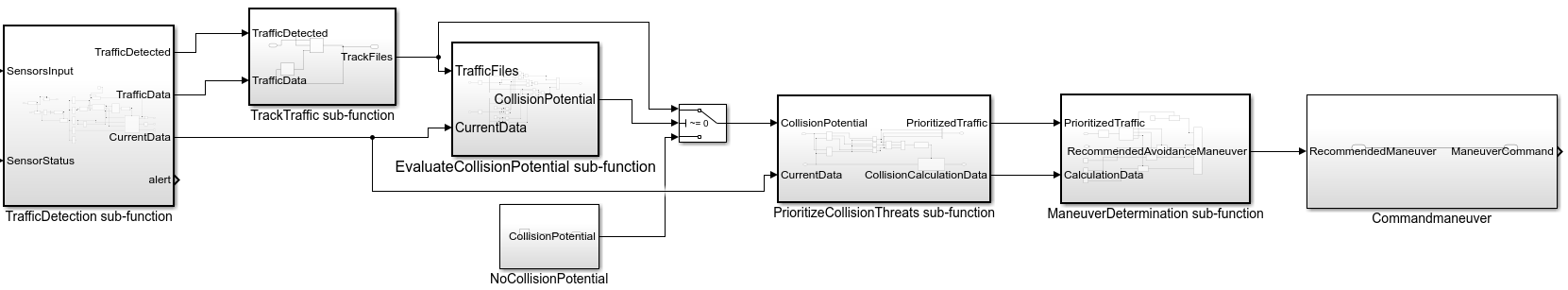}
\caption{Simulink model of the CAS showing subsystems and data flow.}
\label{fig:Simulinkmodel}
\end{figure}

The CAS’s functional behaviour was initially modelled in MATLAB Simulink \cite{matlab2023simulink}, following the specifications outlined in LLRs, DLLRs, and the overall architecture. Simulink models provided a detailed representation of each subsystem, encapsulating their individual behaviours, data flows, and interactions. Figure \ref{fig:Simulinkmodel} illustrates the Simulink model of the CAS, showcasing its modular subsystems and data flows. It highlights how functional components were structured and interlinked to achieve the overall operational objectives of the system.

Within Simulink, tests were run to identify and eliminate any typos, logical errors, or inconsistencies in the model, creating a solid basis for code creation. Once validated, the model was translated to C code with Embedded Coder, which created separate .c and .h files for each subsystem, resulting in modular code that is consistent with the system's architectural design.

Although automated code generation formed the majority of the implementation, manual coding was required to integrate the subsystems into a unified software application. This involved programming the interactions and dependencies between subsystems, such as data interchange, control flow, and module timing. These adjustments supplement the automated programming and provide more flexibility in managing complicated interactions inside the system. The generated code complies with DO-178C standards for safety-critical systems, giving close attention to traceability, modularization, and design descriptions.

To preserve traceability, all manual code adjustments were rigorously documented, linking each modification to its corresponding requirements in the LLRs and architectural design. This comprehensive documentation ensured continued adherence to DO-178C objectives, particularly A5.5 (traceability of source code to design).

\begin{figure}[hbt!]
\centering
\includegraphics[width=0.5\textwidth]{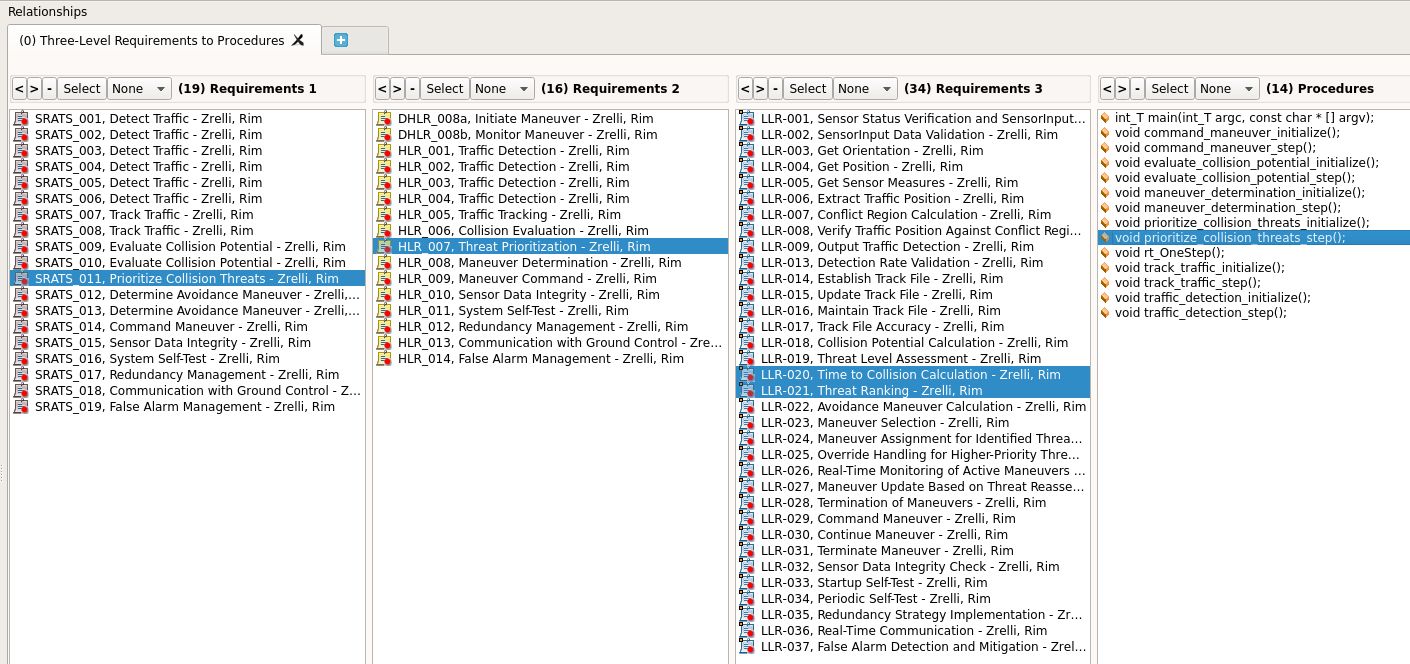}
\caption{TBmanager relationship view showing the traceability between three-requirement levels (SRATS, HLR and LLR) and the source code functions}
\label{fig:TbManagerRelationship}
\end{figure}

To integrate the generated and manually adjusted code into the project’s configuration and verification framework, all code files were imported into TBManager. This integration enabled comprehensive traceability by mapping code functions back to the LLRs and, subsequently, to the original SRATS through HLRs. As depicted in Figure \ref{fig:TbManagerRelationship}, TBManager provides a relational view illustrating the connections between the three levels of requirements and the associated source code procedures.

This traceability structure establishes a clear lineage from high-level system specifications to the implemented code, ensuring that each code function is precisely aligned with the design. It allows for extensive verification, simplifies maintenance tasks, and ensures compliance with DO-178C objectives, particularly A5.5 (traceability of source code to requirements). By maintaining this level of traceability, the project guarantees that the implementation remains consistent with the original design and functional objectives while supporting efficient identification and resolution of potential issues during future updates or audits.

To ensure the correctness and reliability of the code, both static and dynamic code analyses were conducted using the LDRA tool suite.

The generated and manually refined code underwent static analysis to detect potential coding standard violations, evaluate code quality, identify data flow issues, and detect runtime anomalies. LDRA’s TBvision tool was employed to conduct this analysis, generating comprehensive static analysis reports that flagged potential violations and ensured adherence to DO-178C objectives A5.4 (conformity to standards) and A5.6 (accuracy and consistency).  The analysis began by verifying the source code for conformance with MISRA C 2012 \cite{misra2012}, a widely used coding standard for embedded systems. The choice is made as Embedded Coder that generated source code uses this standard.

\begin{figure}[hbt!]
\centering
\includegraphics[width=0.5\textwidth]{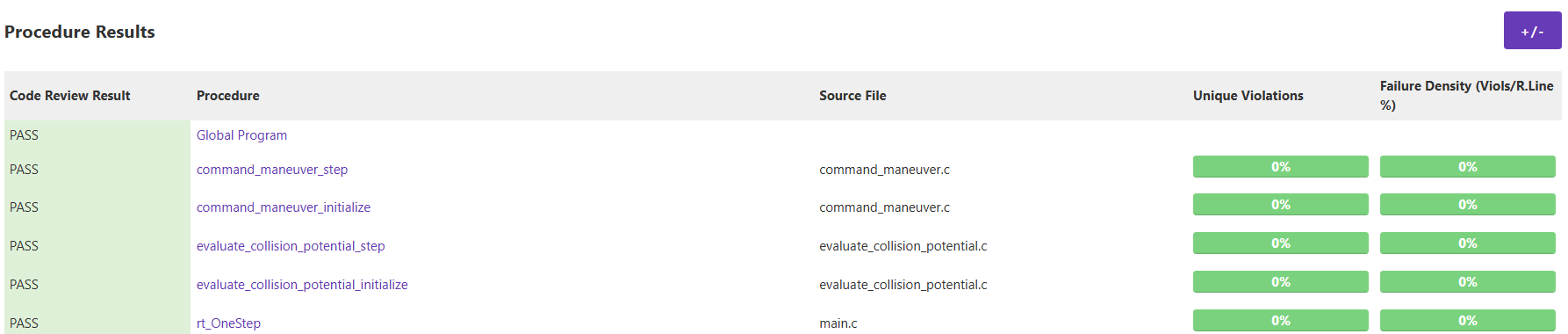}
\caption{Code review report generated by LDRA}
\label{fig:codeReview}
\end{figure}

The static analysis began with an initial scan where TBvision found multiple breaches classified as "required", which must be addressed to comply with the standard. This initial assessment marked the start of an iterative procedure in which violations were systematically addressed and then reanalyzed to assure compliance. During these cycles, the interactive refinement of the code resolved the required violations, allowing for increasing conformity with the standard. In the final revisions, all the violations were treated. Achieving coding standard compliance ensures the source code adheres to a predetermined set of guidelines. Part of the results of the code review are summarized in Figure \ref{fig:codeReview}. Procedures marked as "PASS" indicate that the referred procedures fully adhered to the standard.

Once compliance with coding standards was achieved, the focus shifted to assessing software quality through a structured evaluation of key metrics: clarity, maintainability, and testability. 
\begin{itemize}
    \item \textbf{Clarity}: Ensures the code is easily understandable, facilitating peer reviews, debugging, and future modifications.
    \item \textbf{Maintainability}: Evaluates the ease of implementing changes in response to evolving requirements, ensuring adaptability and resilience over the software lifecycle.
    \item \textbf{Testability}: Measures how readily the code can be tested, which is crucial for ensuring functional accuracy and identifying defects efficiently.
\end{itemize}

\begin{figure}[hbt!]
\centering
\includegraphics[width=0.5\textwidth]{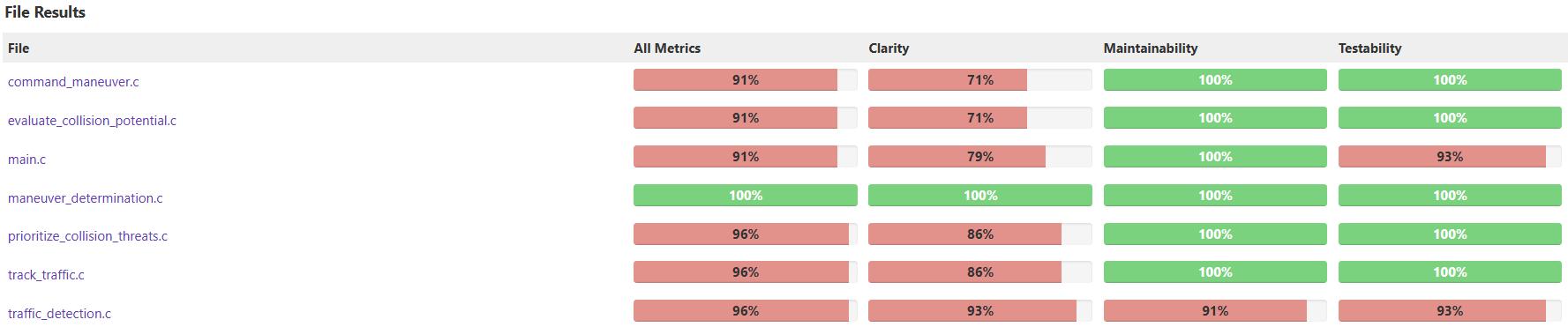}
\caption{Code quality report generated by LDRA}
\label{fig:codeQuality}
\end{figure}

Using LDRA's analysis tools, each metric was thoroughly evaluated, and the results are depicted in Figure \ref{fig:codeQuality}. The analysis revealed that most source files scored exceptionally well in maintainability and testability, indicating a robust design. While the clarity metric showed areas for improvement in specific files, these did not impact compliance or functionality, but they highlight opportunities for further refinement to enhance code readability.

\textbf{Dynamic Code Analysis}: The dynamic analysis of the code concentrates on determining structural coverage, which evaluates the extent to which software code is exercised under certain test conditions. This analysis provides a dynamic metric essential for ensuring the thoroughness of testing. These criteria are matched to the criticality of the code under test, ensuring proportionate efforts to achieve acceptable coverage. Automated tools, such as LDRA, are strongly suggested for this process due to their efficiency and accuracy, particularly in complex systems.

\begin{figure}[hbt!]
\centering
\includegraphics[width=0.5\textwidth]{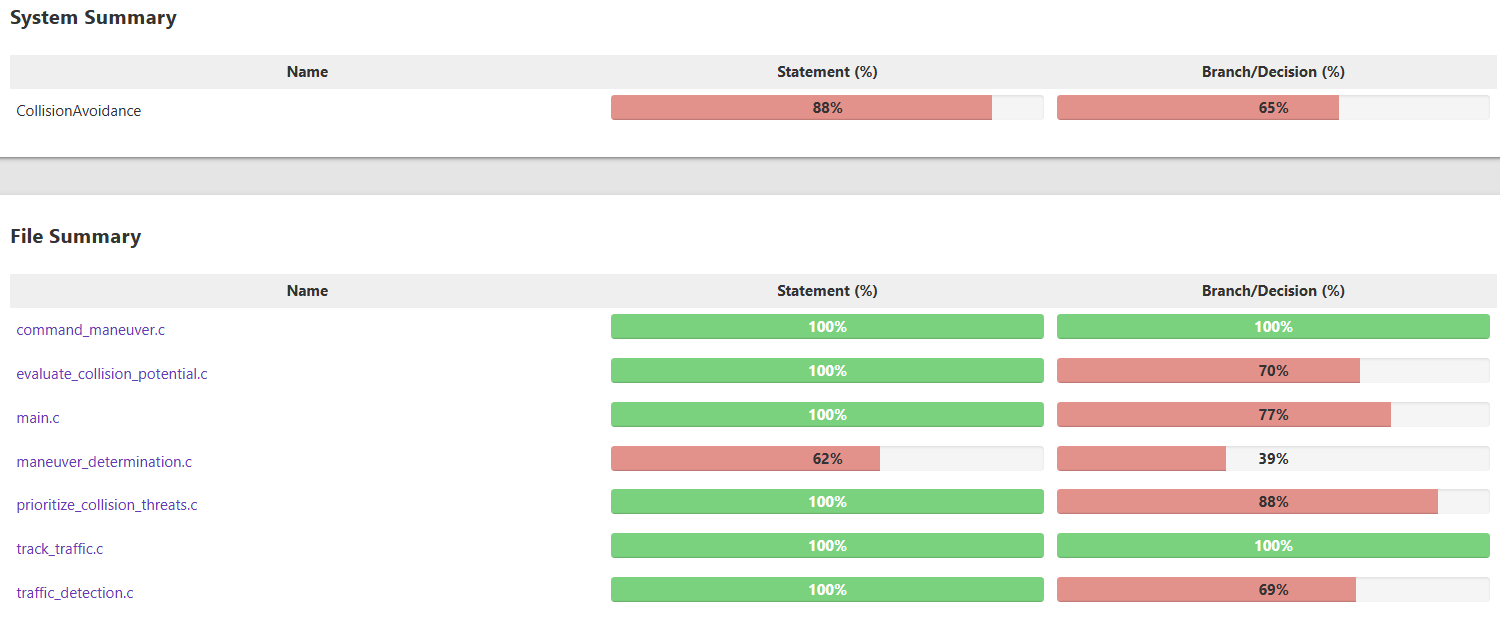}
\caption{Code coverage analysis report generated by LDRA}
\label{fig:codeCoverage}
\end{figure}

The structural coverage analysis was carried out using LDRA's TBvision tool. The process entailed instrumenting the source code, in which LDRA added additional code to monitor execution during testing. We provided simulated inputs to the application execution representing UAV position data and sensor detections, modelling various scenarios of traffic presence or absence. These inputs tested various paths and circumstances within the code. The analysis produced extensive results, as shown in Figure \ref{fig:codeCoverage}, that included percentage coverage for key metrics including Statement and Branch/Decision coverage for each file, as well as specific information about untested code paths.

The evaluation was conducted to meet the requirements of DO-178C Level B software, confirming that the coverage metrics were appropriate for this criticality level. The assessment confirmed if the required structural coverage parameters for this safety level were met. While the procedure needed careful planning and iterative testing to resolve coverage gaps, LDRA's thorough insights helped to improve code quality and ensure compliance with stringent avionics software standards. This emphasizes the role of dynamic analysis in showing software dependability and functional safety in critical systems.

The code coverage results provide useful information about the UAV software system's performance and reliability. While 100\% Statement and Branch/Decision coverage across all files is desirable, the existing results identify areas for improvement and serve as a benchmark. For example, in \textit{maneuver\_determination.c}, which obtained 62\% Statement and 39\% Branch/Decision coverage, the analysis highlighted particular regions where Simulink-generated support functions like \textit{rtIsNaNF} were not exercised during testing. These findings demonstrate the LDRA analysis's effectiveness in identifying potential optimization opportunities, even inside automatically generated code.

The initial results of the code coverage analysis performed using TBvision, with a set of input data, revealed lower-than-expected Statement and Branch/Decision coverage percentages for several files. However, LDRA allows us to exercise the code using test cases run in TBrun, allowing us to refine and expand our testing. We estimate that by using TBrun to run more specialized test cases, we will improve the coverage results for the bulk of the files, resulting in more extensive code validation.

Overall, the results demonstrate the effectiveness of using automated techniques in UAV software development. The coverage attained shows that essential areas of the code were thoroughly tested, with detailed reports providing actionable insights for targeted improvements. Importantly, the procedure validates the idea of employing LDRA to improve the rigour and efficiency of the development process.

Formal verification was employed during the coding phase to ensure that the CAS code adhered to critical safety properties, aligning with DO-178C objectives for verifiability (A5.3) and accuracy (A5.6). Bounded Model Checking \cite{biere2021bounded} was utilized to validate the correctness of code segments by modelling the code logic and verifying it against pre-defined safety properties. This approach provided mathematical assurance of the system’s logical soundness and operational reliability.

To this end, the Efficient SMT-Based Bounded Model Checker (ESBMC) was chosen for its compatibility with the C programming language and its capability to simulate program execution across all possible input scenarios. ESBMC is designed to detect a wide array of potential programming issues, including out-of-bounds array accesses, null or invalid pointer dereferences, memory alignment violations, integer overflows, and floating-point anomalies (e.g., NaNs and division by zero). Additionally, it aids in identifying memory leaks, a critical factor in ensuring code robustness and compliance with DO-178C standards for resilience and dependability.

To use ESBMC for verifying the correctness of our code, the first step is to specify input handling with nondeterministic values that ESBMC can explore during bounded model checking. Specifically, sensor values inputs were defined as \textit{nondet\_double()}, ensuring that the model checker can test a wide range of potential values. Furthermore, $\_\_ESBMC\_assume$ is used to apply constraints on the sensor inputs, limiting the values to a suitable range depending on the system's predicted behaviour. This is important because ESBMC works by investigating every possible combination of inputs, and it needs deterministic inputs to verify every possible program execution path.

\begin{figure}[hbt!]
\centering
\includegraphics[width=0.5\textwidth]{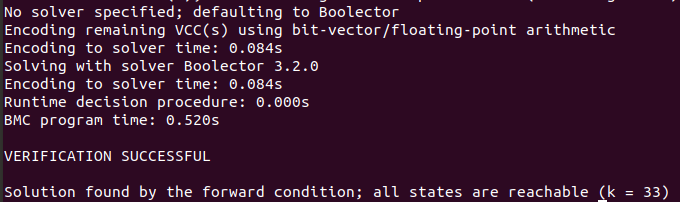}
\caption{Bounded Model Checking results using ESBMC for CAS code}
\label{fig:bmc_results}
\end{figure}

The verification process leveraged ESBMC’s ability to analyze and prove that all states within the program were reachable under the forward condition. This comprehensive analysis was exemplified by the successful verification output shown in Figure \ref{fig:bmc_results}, which demonstrates that the CAS code satisfied all specified properties after exhaustive exploration of execution paths, with a maximum bounded state depth of $k=33$.

The final step in the coding phase involved a thorough review and automated testing to ensure that the code met all safety and performance standards. LDRA was employed to conduct automated testing and generate test reports, ensuring the code's robustness and compliance.

\begin{figure}[hbt!]
\centering
\includegraphics[width=0.5\textwidth]{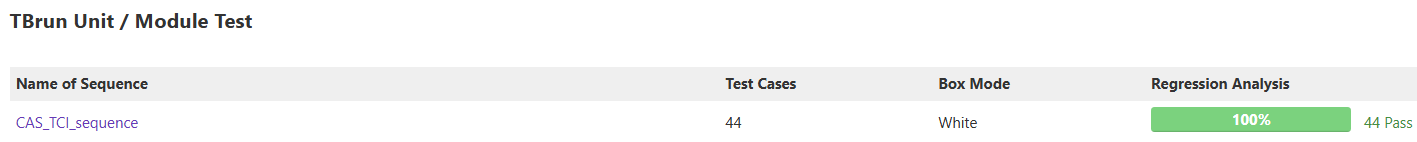}
\caption{Tbrun Unit / Module Test report}
\label{fig:tbrunTest}
\end{figure}

\begin{itemize} 
    \item Code Review: The generated and modified code was systematically reviewed by experts to ensure that it adhered to the system’s functional and safety requirements. The review process focused on verifying that the code conformed to the architectural models and complied with DO-178C objective A5.1 (compliance with LLRs). 
    \item Automated Testing: Automated test cases were performed using TBrun, where a sequence of test cases was designed and executed to test the source code thoroughly. TBrun enables the execution of procedure calls using the specified test data, guaranteeing that the source code works as intended. The test data was defined using test cases drawn from HLRs and LLRs, ensuring complete coverage of the system's requirements. In TBrun, a white-box analysis was performed, with an emphasis on a thorough structural evaluation of the code under test. This analysis used Coverage Analysis to determine the usefulness of the test data in running the unit tests and guaranteeing the code's functionality.

    A total of 44 test cases were run, with the results indicating that 100\% of them passed properly, as depicted in Figure \ref{fig:tbrunTest}. A manual review was also performed to confirm that the expected outputs were reached, adding another layer of validation to the tests' accuracy and completeness. This comprehensive method not only evaluated the software's conformance to functional requirements but also helped to ensure the UAV system's overall reliability and safety. These tests ensured that each component functioned correctly and that the code complied with DO-178C objectives A5.1 (compliance with LLRs) and A5.2 (compliance with software architecture).
\end{itemize}

\begin{figure}[hbt]
\centering
\includegraphics[width=0.5\textwidth]{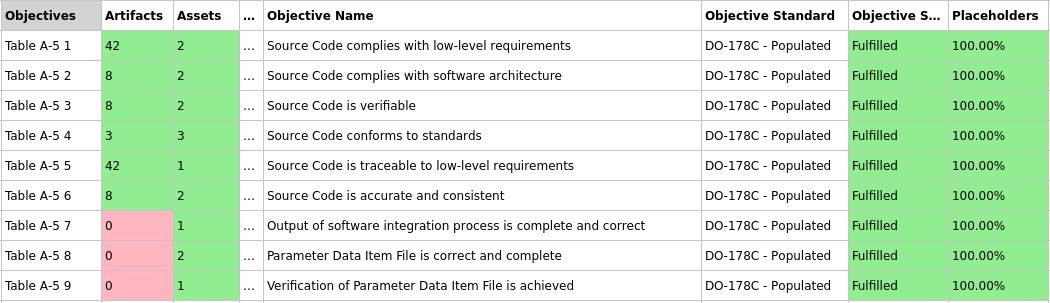}
\caption{DO-178C objectives for verification of outputs of the software coding process in TBmanager}
\label{fig:ObjectivesA5}
\end{figure}

By the end of the coding phase, the CAS source code was fully generated, verified, and validated through a combination of formal methods, manual reviews, and automated testing. Figure \ref{fig:ObjectivesA5} provides a summary of how the code satisfied DO-178C coding objectives. This rigorous process ensured the code adhered to all required safety and performance standards, forming a robust and compliant foundation for the subsequent Software Integration phase.

\section{Summary of Implementation Outcomes}
The implementation of the proposed methodology for developing a DAL-B CAS for UAVs allowed significant outcomes, demonstrating both the strengths and practical challenges of integrating formal methods with automated verification tools for DO-178C compliance.
\subsection*{Key Achievements}
\begin{itemize}
    \item Rigorous requirement validation: Formal methods, notably the Alloy analyzer, were successfully leveraged to specify and verify the CAS system requirements (SRATS, HLRs, DHLRs). Early application of formal specification led to the identification and resolution of ambiguities and logical inconsistencies, particularly in threat detection and prioritization requirements. Iterative refinement and peer reviews produced requirements artifacts that were both verifiable and traceable, fully aligned with DO-178C objectives.

    \item Design verification and traceability: The architecture and LLRs were defined using a modular approach, with each subsystem (e.g., Traffic Detection, Threat Prioritization, Maneuver Determination) formally specified and verified using Spin and Alloy. Automated tools (TBmanager) provided robust bidirectional traceability from requirements through to design artifacts, ensuring that all design elements remained linked to their originating requirements. Formal model checking confirmed the absence of deadlocks and the satisfaction of critical safety properties.

    \item Automated and formally-assured coding: Simulink Embedded Coder was employed to automate code generation from verified models, reducing the risk of manual errors and ensuring traceability to design artifacts. The generated code underwent both static and dynamic analysis using the LDRA tool suite, which confirmed compliance with coding standards and design intent. Formal verification (bounded model checking) was applied to safety-critical code sections, providing mathematical assurance of correctness.

    \item Verification and validation evidence: Automated testing and objective summary reports generated by LDRA tools, along with comprehensive traceability matrices, documented the status of DO-178C objectives across all implemented phases. These artifacts provide concrete evidence of compliance, supporting certification activities and peer review processes.
    
\end{itemize}
\subsection*{Limitation}
While the requirements, design, and coding phases were fully realized and documented, the integration phase was not comprehensively implemented. Automated integration testing and end-to-end system verification, as prescribed in the methodology, remain areas for future development. As a result, full system-level verification, including Parameter Data Item (PDI) generation and integrated deployment validation, is planned for subsequent work.

\subsection*{Lessons Learned}
The case study demonstrated that integrating formal methods with automated tool suites significantly improves the rigor and efficiency of the requirements, design, and coding activities for safety-critical UAV software. Early error detection, automated traceability, and objective evidence generation streamlined the certification process and mitigated many manual effort risks. 

Overall, the outcomes validate the feasibility and benefits of the proposed methodology, while also identifying practical gaps that will inform ongoing and future work towards full DO-178C compliance for autonomous UAV systems.

\section*{ACKNOWLEDGMENTS}

This research was supported by the Consortium for Aerospace Research and Innovation in Quebec (CRIAQ) and by the
Fonds of Mitacs under grant IT19246 and grant IT30530.


\appendices
\onecolumn
\section{System Requirements Allocated to Software (SRATS)} \label{App:srats}
\begin{longtable}{p{2cm} p{13.5cm}}
\caption{SRATS List}\\

SRATS\_001 & \textit{Detect Traffic}: The Collision Avoidance System shall detect traffic within its surveillance volume. \newline \textit{Rationale}: Effective traffic detection is the first step in collision avoidance, ensuring that potential threats are identified early. \newline \textit{Category}: Functional \newline \textit{Traceability}: Developed into HLR\_001
\\
SRATS\_002 & \textit{Detect Traffic}: The Collision Avoidance System shall calculate its surveillance volume depending on all the following conditions: 
\begin{itemize}
    \item detection range;
    \item azimuth field of regard;
    \item elevation field of regard.
\end{itemize}
\textit{Rationale}: Defines the area within which traffic detection is performed.
\newline \textit{Category}: Functional \newline \textit{Traceability}: Developed into HLR\_001
\\
SRATS\_003 & \textit{Detect Traffic}: The Collision Avoidance System shall detect cooperative traffic at a range of at least 20 nautical miles. \newline \textit{Rationale}: Ensures the system can detect cooperative traffic, which is essential for collision avoidance. \newline \textit{Category}: Functional \newline \textit{Traceability}: Developed into HLR\_002
\\
SRATS\_004 & \textit{Detect Traffic}: The Collision Avoidance System shall detect cooperative traffic within an azimuth FOR of at least +/-110° referenced from the flight path of the UA.  \newline \textit{Rationale}: Ensures wide horizontal coverage for detecting traffic. \newline \textit{Category}: Functional \newline \textit{Traceability}: Developed into HLR\_003
\\
SRATS\_005 & \textit{Detect Traffic}: The Collision Avoidance System shall detect cooperative traffic within an elevation FOR of at least +/-15° referenced from the flight path of the UA.   \newline \textit{Rationale}: Ensures vertical coverage for detecting traffic. \newline \textit{Category}: Functional \newline \textit{Traceability}: Developed into HLR\_003
\\
SRATS\_006 & \textit{Detect Traffic}: The average Collision Avoidance System detection rate shall be equal to or greater than 1.0 hertz.  \newline \textit{Rationale}:A sufficient detection rate is necessary to ensure timely updates and responses to detected traffic. \newline \textit{Category}: Functional \newline \textit{Traceability}: Developed into HLR\_004
\\
SRATS\_007 & \textit{Track Traffic}: The Collision Avoidance System shall track the detected traffic. \newline \textit{Rationale}: Accurate tracking of traffic allows for continuous monitoring and assessment of potential collision threats. \newline \textit{Category}: Functional \newline \textit{Traceability}: Developed into HLR\_005
\\
SRATS\_008 & \textit{Track Traffic}: The Collision Avoidance System shall track cooperative traffic within an azimuth FOR of at least +/-110° and elevation FOR of at least +/-15° referenced from the flight path of the UA. \newline \textit{Category}: Functional \newline \textit{Traceability}: Developed into HLR\_005
\\
SRATS\_009 & \textit{Evaluate Collision Potential}: The Collision Avoidance System shall evaluate the potential for collision with each tracked traffic element, including the assessment of existing collision threats.  \newline \textit{Rationale}: Evaluating collision potential helps in identifying imminent threats and preparing the system to take appropriate actions. \newline \textit{Category}: Functional \newline \textit{Traceability}: Developed into HLR\_006
\\
SRATS\_010 & \textit{Evaluate Collision Potential}: The Collision Avoidance System shall continuously determine if any detected traffic elements pose a collision threat to the vehicle. \newline \textit{Category}: Functional \newline \textit{Traceability}: Developed into HLR\_006
\\
SRATS\_011 & \textit{Prioritize Collision Threats}: The Collision Avoidance System shall prioritize the traffic posing a collision threat.  \newline \textit{Rationale}: Prioritizing threats ensures that the most immediate and dangerous threats are addressed first, optimizing response time and effectiveness. \newline \textit{Category}: Functional \newline \textit{Traceability}: Developed into HLR\_007
\\
SRATS\_012 & \textit{Determine Avoidance Maneuver}: The Collision Avoidance System shall determine an avoidance maneuver that prevents a collision. \newline \textit{Rationale}: Determining the correct avoidance maneuver is essential to ensure the safety of the UAV and other airspace users. \newline \textit{Category}: Functional \newline \textit{Traceability}: Developed into HLR\_008
\\
SRATS\_013 & \textit{Determine Avoidance Maneuver}: The Collision Avoidance System shall revise the maneuver recommendation when other aircraft are simultaneously maneuvering. \newline \textit{Category}: Functional \newline \textit{Traceability}: Developed into HLR\_008
\\
SRATS\_014 & \textit{Command Maneuver}: The Collision Avoidance System shall command an appropriate avoidance maneuver.  \newline \textit{Rationale}: Executing the correct maneuver ensures the UAV can avoid collisions effectively. \newline \textit{Category}: Functional \newline \textit{Traceability}: Developed into HLR\_009
\\
SRATS\_015 & \textit{Sensor Data Integrity}: The Collision Avoidance System shall verify the integrity and accuracy of the sensor data before using it for detection and tracking. \newline \textit{Rationale}: Ensuring that the sensor data is reliable and accurate is critical for all subsequent functions of the Collision Avoidance System. \newline \textit{Category}: Performance \newline \textit{Traceability}: Developed into HLR\_010
\\
SRATS\_016 & \textit{System Self-Test}: The Collision Avoidance System shall perform self-tests on startup and periodically during operation to ensure all components are functioning correctly.  \newline \textit{Rationale}: Regular self-tests help in identifying and mitigating failures that could compromise the system's ability to detect and avoid collisions. \newline \textit{Category}: Safety \newline \textit{Traceability}: Developed into HLR\_011
\\
SRATS\_017 & \textit{Redundancy Management}: The Collision Avoidance System shall have redundancy management capabilities to handle failures in primary detection and tracking components.  \newline \textit{Rationale}: Redundancy ensures that the system remains operational even if some components fail, thus maintaining safety. \newline \textit{Category}: Safety \newline \textit{Traceability}: Developed into HLR\_012
\\
SRATS\_018 & \textit{Communication with Ground Control}: The Collision Avoidance System shall communicate status and alerts to the ground control station in real-time. \newline \textit{Rationale}: Real-time communication with ground control allows for human intervention when necessary and provides situational awareness to operators. \newline \textit{Category}: Non-Functional \newline \textit{Traceability}: Developed into HLR\_013
\\
SRATS\_019 & \textit{False Alarm Management}: The Collision Avoidance System shall minimize false alarms to prevent unnecessary maneuvers.  \newline \textit{Rationale}: Reducing false alarms ensures that the system does not execute unnecessary maneuvers, which could be disruptive or hazardous. \newline \textit{Category}: Performance \newline \textit{Traceability}: Developed into HLR\_014
\\
\label{tab:SRATS}
\end{longtable}

\section{High-Level Requirements (HLRs)} \label{App:hlr}
\begin{longtable}{p{2cm} p{13.5cm}}
\caption{HLRs List}\\
HLR\_001 & \textit{Traffic Detection}: The Collision Avoidance System shall detect traffic within its surveillance volume. Note: The surveillance volume is defined by three performance characteristics of the sensor: detection range, azimuth field of regard, and elevation field of regard.  \newline \textit{Rationale}: Ensuring that traffic within the defined surveillance volume is detected is critical for initiating subsequent collision avoidance processes. \newline \textit{Traceability}: Developed from SRATS\_001 and SRATS\_002. Developed into LLR-001, LLR-002, LLR-003, LLR-004, LLR-005, LLR-006, LLR-007, LLR-008, and LLR-009.
\\

HLR\_002 & \textit{Traffic Detection}: The Collision Avoidance System shall detect cooperative traffic at a range of at least 20 nautical miles.   \newline \textit{Rationale}: Ensuring the detection of cooperative traffic within a significant range is essential for collision avoidance. \newline \textit{Traceability}: Developed from SRATS\_003. Developed into LLR-0010.
\\
HLR\_003 & \textit{Traffic Detection}: The Collision Avoidance System shall detect cooperative traffic within an azimuth field of regard of at least ±110° and an elevation field of regard of at least ±15°, referenced from the flight path of the UAV.  \newline \textit{Rationale}: Detecting traffic within these fields of regard ensures comprehensive surveillance around the UAV. \newline \textit{Traceability}: Developed from SRATS\_004 and SRATS\_005. Developed into LLR-0011 and LLR-0012.
\\
HLR\_004 & \textit{Traffic Detection}: The average Collision Avoidance System detection rate shall be equal to or greater than 1.0 hertz.  \newline \textit{Rationale}: A sufficient detection rate is necessary to ensure timely updates and responses to detected traffic. \newline \textit{Traceability}: Developed from SRATS\_006. Developed into LLR-0013.
\\
HLR\_005 & \textit{Traffic Tracking}: The Collision Avoidance System shall track the detected traffic. Note: The track is established when a state estimate is developed with sufficient confidence, and this estimate includes the traffic element’s position and velocity vector.  \newline \textit{Rationale}: Accurate tracking of detected traffic is essential for evaluating potential collision threats and determining appropriate avoidance maneuvers. \newline \textit{Traceability}: Developed from SRATS\_007 and SRATS\_008. Developed into LLR-0014, LLR-0015, LLR-0016, and LLR-0017.
\\
HLR\_006 & \textit{Collision Evaluation}: The Collision Avoidance System shall evaluate the potential for collision with each traffic element being tracked. This evaluation includes assessing existing collision threats.  \newline \textit{Rationale}: Evaluating collision potential is necessary to identify imminent threats and prepare the system for appropriate responses. \newline \textit{Traceability}: Developed from SRATS\_009 and SRATS\_010. Developed into LLR-0018 and LLR-0019.
\\
HLR\_007 & \textit{Threat Prioritization}: The Collision Avoidance System shall prioritize the traffic posing a collision threat. Prioritization is based on ranking the time to collision of the identified threats. \newline \textit{Rationale}: Prioritizing threats ensures that the most immediate and dangerous threats are addressed first, optimizing the system's response time and effectiveness. \newline \textit{Traceability}: Developed from SRATS\_011. Developed into LLR-0020 and LLR-0021.
\\
HLR\_008 & \textit{Maneuver Determination}: The Collision Avoidance System shall autonomously determine an avoidance maneuver that prevents a collision.  \newline \textit{Rationale}: Autonomous determination of avoidance maneuvers is critical to ensure timely and effective responses to imminent collision threats. \newline \textit{Traceability}: Developed from SRATS\_012 and SRATS\_013. Developed into LLR-0022 and LLR-0023.
\\
HLR\_009 & \textit{Maneuver Command}: The Collision Avoidance System shall command an appropriate avoidance maneuver.  Note: The commanded maneuver can include initiating a new maneuver, continuing an ongoing maneuver, or terminating an avoidance maneuver if a collision threat no longer exists. \newline \textit{Rationale}: Executing the appropriate maneuver ensures the avoidance of collisions based on real-time evaluations of traffic and threats. \newline \textit{Traceability}: Developed from SRATS\_014. Developed into LLR-0024, LLR-0025, and LLR-0026.
\\
HLR\_010 & \textit{Sensor Data Integrity}: The CAS shall verify the integrity and accuracy of the sensor data before using it for detection and tracking. \newline \textit{Rationale}: Ensuring that the sensor data is reliable and accurate is critical for all subsequent functions of the CAS. \newline \textit{Traceability}: Developed from SRATS\_015. Developed into LLR-0027.
\\
HLR\_011 & \textit{System Self-Test}: The CAS shall perform self-tests on startup and periodically during operation to ensure all components are functioning correctly. \newline \textit{Rationale}: Regular self-tests help in identifying and mitigating failures that could compromise the system's ability to detect and avoid collisions. \newline \textit{Traceability}: Developed from SRATS\_016. Developed into LLR-0028 and LLR-0029.
\\
HLR\_012 & \textit{Redundancy Management}: The CAS shall have redundancy management capabilities to handle failures in primary detection and tracking components.  \newline \textit{Rationale}: Redundancy ensures that the system remains operational even if some components fail, thus maintaining safety. \newline \textit{Traceability}: Developed from SRATS\_017. Developed into LLR-0030.
\\
HLR\_013 & \textit{Communication with Ground Control}: The CAS shall communicate status and alerts to the ground control station in real-time. \newline \textit{Rationale}: Real-time communication with ground control allows for human intervention when necessary and provides situational awareness to operators. \newline \textit{Traceability}: Developed from SRATS\_018. Developed into LLR-0031
\\
HLR\_014 & \textit{False Alarm Management}: The CAS shall minimize false alarms to prevent unnecessary maneuvers. \newline \textit{Rationale}: Reducing false alarms ensures that the system does not execute unnecessary maneuvers, which could be disruptive or hazardous. \newline \textit{Traceability}: Developed from SRATS\_019. Developed into LLR-0032.
\\
\label{tab:HLR}
\end{longtable}

\section{CAS Architecture \& Low-Level Requirements (LLRs)} \label{App:llr}
\subsection{CAS Architecure}

\begin{figure}[H]
\centering
\includegraphics[width=1\textwidth]{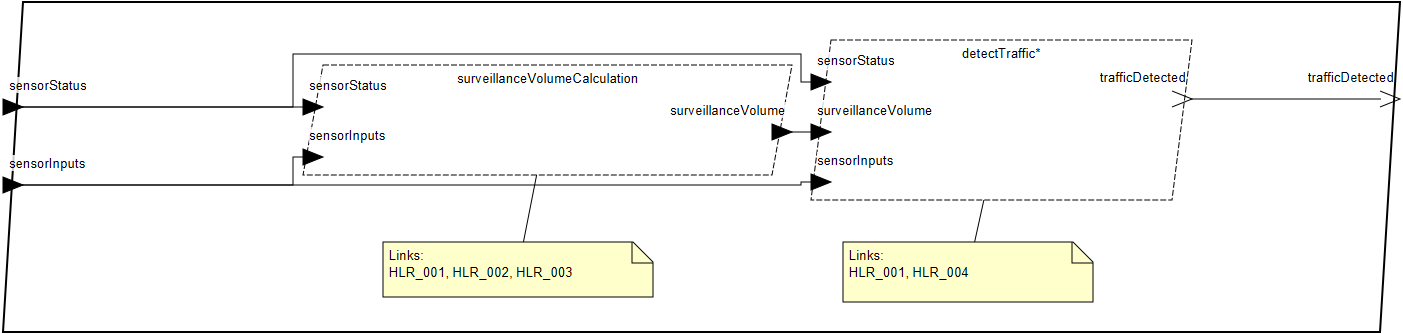}
\caption{AADL model of the traffic detection subsystem}
\label{fig:TrafficDetecAADL}
\end{figure}

\begin{figure}[H]
\centering
\includegraphics[width=1\textwidth]{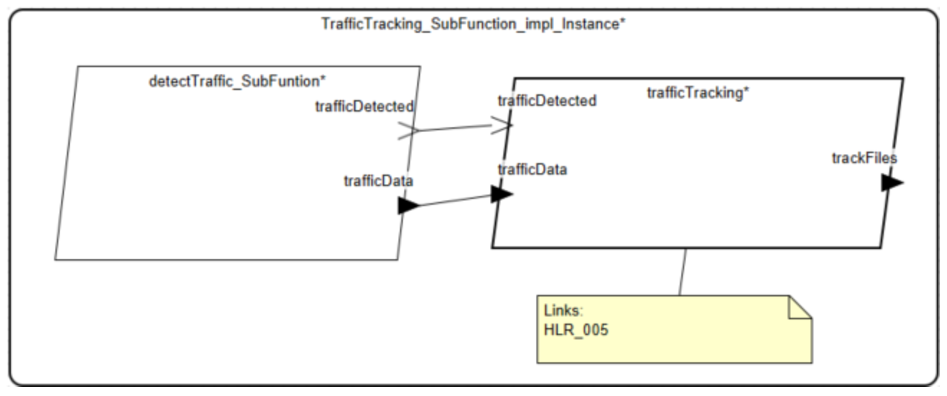}
\caption{AADL model of the traffic tracking subsystem}
\label{fig:TrafficTrackingAADL}
\end{figure}

\begin{figure}[H]
\centering
\includegraphics[width=1\textwidth]{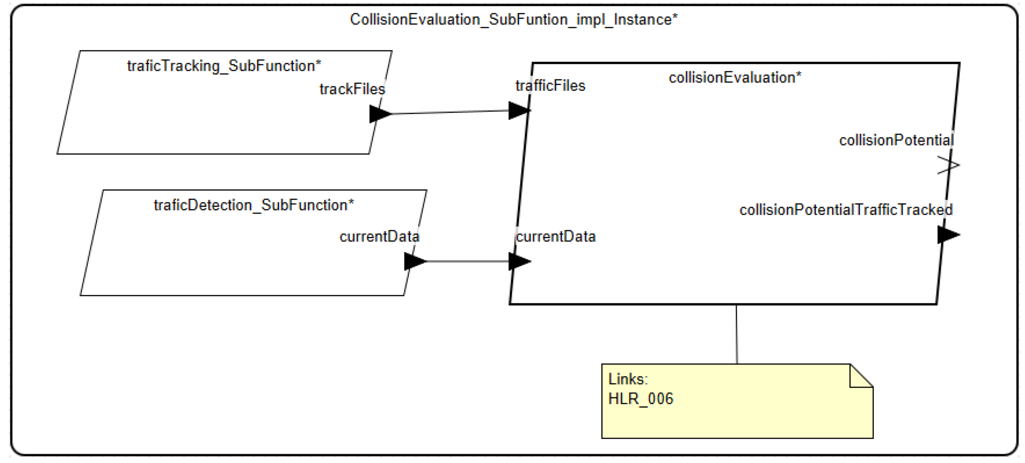}
\caption{AADL model of the collision evaluation subsystem}
\label{fig:ColEvalAADL}
\end{figure}

\begin{figure}[H]
\centering
\includegraphics[width=1\textwidth]{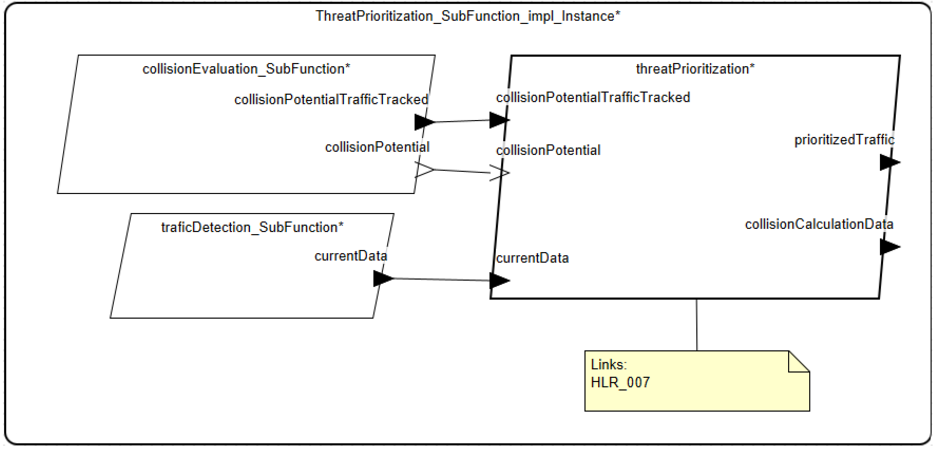}
\caption{AADL model of the threat prioritization subsystem}
\label{fig:ThreatPrioAADL}
\end{figure}

\begin{figure}[H]
\centering
\includegraphics[width=1\textwidth]{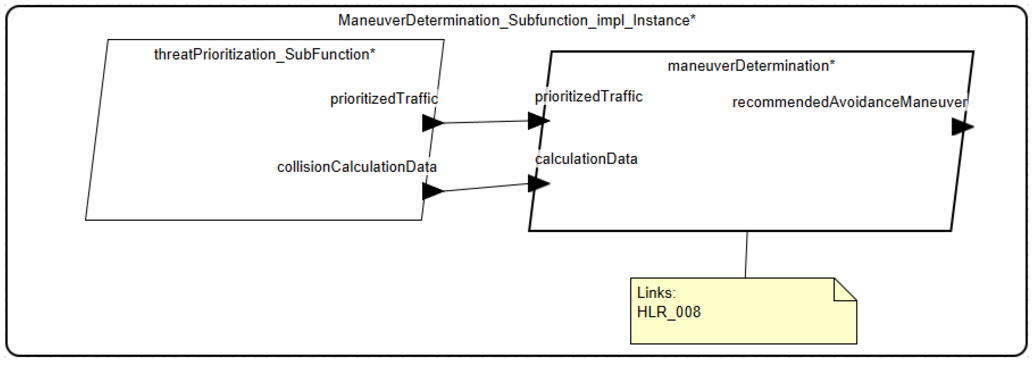}
\caption{AADL model of the maneuver determination subsystem}
\label{fig:ManDetAADL}
\end{figure}

\begin{figure}[H]
\centering
\includegraphics[width=1\textwidth]{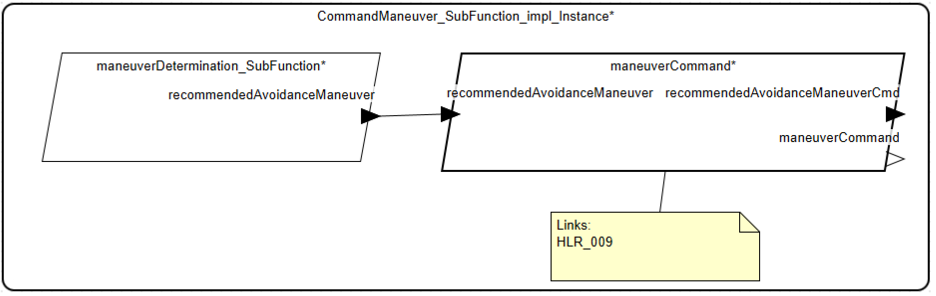}
\caption{AADL model of the maneuver command subsystem}
\label{fig:ManComAADL}
\end{figure}

\subsection{LLRs}
\begin{longtable}{p{2cm} p{13.5cm}}
\caption{LLRs List}\\
LLR\_001 & \textit{Sensor Status Verification and SensorInput Retrieval}: When the SensorStatus is True, the TrafficDetection shall retrieve the SensorInput. Otherwise, the TrafficDetection shall call the Alert function. \newline \textit{Traceability}: Developed from HLR-001.
\\

LLR\_002 & \textit{SensorInput Data Validation}: When SensorInput is received, TrafficDetection shall validate the data as follows: - DetectionRange value is between 0.2 and 3000 meters. -AzimuthFOR value is between -110 and 110 degrees. - ElevationFOR value is between -15 and 15 degrees. \newline \textit{Traceability}: Developed from HLR-001.
\\
LLR\_003 & \textit{Get Orientation}: When SensorInput data is validated, TrafficDetection shall get the UAV orientation using the OrientationData from SensorInput. \newline \textit{Traceability}: Developed from HLR-001.
\\
LLR\_004 & \textit{Get Position}: When SensorInput data is validated, TrafficDetection shall get the current position using the PositionData from SensorInput. \newline \textit{Traceability}: Developed from HLR-001.
\\
LLR\_005 & \textit{Get Sensor Measures}: When SensorInput data is validated, TrafficDetection shall get the UAV sensor measures using the SensorMeasures from SensorInput. \newline \textit{Traceability}: Developed from HLR-001.
\\
LLR\_006 & \textit{Extract Traffic Position}: When the SensorMeasure is not empty, TrafficDetection shall calculate the TrafficPosition using the SensorMeasure and the CurrentPosition. \newline \textit{Traceability}: Developed from HLR-001.
\\
LLR\_007 & \textit{Conflict Region Calculation}: TrafficDetection shall calculate the ConflictRegion using the CurrentPosition, HorizontalRadius, and VerticalMeasure. \newline \textit{Traceability}: Developed from HLR-001.
\\
LLR\_008 & \textit{Verify Traffic Position Against Conflict Region}: TrafficDetection shall compare the TrafficPosition with the ConflictRegion and do the following: When TrafficPosition is in the ConflictRegion, set PotentialTraffic to True. Otherwise, set PotentialTraffic to False. \newline \textit{Traceability}: Developed from HLR-001.
\\
LLR\_009 & \textit{Output Traffic Detection}: When PotentialTraffic is True, TrafficDetection shall output TrafficDetected, TrafficPosition, and CurrentPosition. \newline \textit{Traceability}: Developed from HLR-001.
\\
LLR\_010 & \textit{Detection Range Validation}: TrafficDetection shall validate that the detection range for cooperative traffic is at least 20 nautical miles. \newline \textit{Traceability}: Developed from HLR-002.
\\
LLR\_011 & \textit{Azimuth FOR Validation}: TrafficDetection shall validate that the azimuth field of regard for cooperative traffic detection is at least ±110°. \newline \textit{Traceability}: Developed from HLR-003.
\\
LLR\_012 & \textit{Elevation FOR Validation}: TrafficDetection shall validate that the elevation field of regard for cooperative traffic detection is at least ±15°. \newline \textit{Traceability}: Developed from HLR-003.
\\
LLR\_013 & \textit{Detection Rate Validation}: TrafficDetection shall ensure that the detection rate is equal to or greater than 1.0 hertz. \newline \textit{Traceability}: Developed from HLR-004.
\\
LLR\_014 & \textit{Establish Track File}: When TrafficDetected is True, the TrafficTracking shall establish a track file for the detected traffic, including Position and VelocityVector. \newline \textit{Traceability}: Developed from HLR-005.
\\
LLR\_015 & \textit{Update Track File}: The TrafficTracking shall update the track file for each detected traffic element at a rate of at least X updates per second. \newline \textit{Traceability}: Developed from HLR-005.
\\
LLR\_016 & \textit{Maintain Track File}: The TrafficTracking shall maintain the track of a detected traffic element until it is no longer detected or poses no threat. \newline \textit{Traceability}: Developed from HLR-005.
\\
LLR\_017 & \textit{Track File Accuracy}: The TrafficTracking shall ensure that the Position and VelocityVector in the track file have an accuracy within Y meters and Z meters/second respectively. \newline \textit{Traceability}: Developed from HLR-005.
\\
LLR\_018 & \textit{Collision Potential Calculation}: The CollisionEvaluation shall calculate the collision potential for each tracked traffic element using Position, VelocityVector, and TrajectoryData. \newline \textit{Traceability}: Developed from HLR-006.
\\
LLR\_019 & \textit{Threat Level Assessment}: The CollisionEvaluation shall assess the threat level of each traffic element based on the calculated collision potential. \newline \textit{Traceability}: Developed from HLR-006.
\\
LLR\_020 & \textit{Time to Collision Calculation}: The ThreatPrioritization shall calculate the TimeToCollision (TTC) for each tracked traffic element. \newline \textit{Traceability}: Developed from HLR-007.
\\
LLR\_021 & \textit{Threat Ranking}: The ThreatPrioritization shall rank traffic elements based on their TimeToCollision, prioritizing the ones with the shortest TTC. \newline \textit{Traceability}: Developed from HLR-007.
\\
LLR\_022 & \textit{Avoidance Maneuver Calculation}: The ManeuverDetermination shall calculate an avoidance maneuver for each identified collision threat. \newline \textit{Traceability}: Developed from HLR-008.
\\
LLR\_023 & \textit{Maneuver Selection}: The ManeuverDetermination shall select the most appropriate avoidance maneuver from the calculated options. \newline \textit{Traceability}: Developed from HLR-008.
\\
LLR\_08a-01 & \textit{Maneuver Assignment for Identified Threats}: The ManeuverDetermination function shall assign a specific avoidance maneuver to each identified collision threat unless a higher-priority threat takes precedence. \newline \textit{Traceability}: Developed from DHLR-008a.
\\
LLR\_08a-02 & \textit{Override Handling for Higher-Priority Threats}: If a higher-priority threat is detected after an initial maneuver is assigned, the CAS shall override the initial maneuver and activate the higher-priority maneuver. \newline \textit{Traceability}: Developed from DHLR-008a.
\\
LLR\_024 & \textit{Command Maneuver}: The ManeuverCommand function shall issue a command to initiate the selected avoidance maneuver. \newline \textit{Traceability}: Developed from HLR-009.
\\
LLR\_025 & \textit{Continue Maneuver}: The ManeuverCommand function shall issue a command to continue an ongoing avoidance maneuver if the threat persists. \newline \textit{Traceability}: Developed from HLR-009.
\\
LLR\_026 & \textit{Terminate Maneuver}: The ManeuverCommand function shall issue a command to terminate an ongoing avoidance maneuver if the threat no longer exists. \newline \textit{Traceability}: Developed from HLR-009.
\\
LLR\_008b-01 & \textit{Real-Time Monitoring of Active Maneuvers}: The CAS shall monitor each active maneuver in real-time, evaluating whether the associated threat still exists or if new threats emerge. \newline \textit{Traceability}: Developed from DHLR-008b.
\\
LLR\_008b-02 & \textit{Maneuver Update Based on Threat Reassessment}: If the threat assessment changes, the CAS shall update the maneuver accordingly. This update can involve adjusting the existing maneuver or initiating a new one. \newline \textit{Traceability}: Developed from DHLR-008b.
\\
LLR\_008b-03 & \textit{Termination of Maneuvers}: The CAS shall terminate an active maneuver when the associated threat is no longer detected or when the maneuver is superseded by a higher-priority action. \newline \textit{Traceability}: Developed from DHLR-008b.
\\
LLR\_027 & \textit{Sensor Data Integrity Check}: The CAS shall perform integrity checks on sensor data before using it for detection and tracking. \newline \textit{Traceability}: Developed from HLR-010.
\\
LLR\_028 & \textit{Startup Self-Test}: The CAS shall perform a self-test on startup to ensure all components are functioning correctly. \newline \textit{Traceability}: Developed from HLR-011.
\\
LLR\_029 & \textit{Periodic Self-Test}: The CAS shall perform self-tests periodically during operation to ensure ongoing functionality of all components. \newline \textit{Traceability}: Developed from HLR-011.
\\
LLR\_030 & \textit{Redundancy Strategy Implementation}: The CAS shall implement redundancy strategies to handle failures in primary detection and tracking components. \newline \textit{Traceability}: Developed from HLR-012.
\\
LLR\_031 & \textit{Real-Time Communication}: The CAS shall transmit status and alerts to the ground control station in real-time. \newline \textit{Traceability}: Developed from HLR-013.
\\
LLR\_032 & \textit{False Alarm Detection and Mitigation}: The CAS shall implement false alarm detection and mitigation strategies to minimize unnecessary maneuvers.  \newline \textit{Traceability}: Developed from HLR-014.
\\
\label{tab:LLR}
\end{longtable}

\section{Failure Mode and Effects Analysis (FMEA)}\label{App:fmea}

\begin{longtable}{|p{2.5cm}|p{2.5cm}|p{3cm}|p{1.5cm}|p{5.5cm}|} \caption{FMEA Report for Collision Avoidance System} \label{tab:FMEA_CAS}\\ 
\hline 
\rowcolor[gray]{0.8} \textbf{Software Element} & \textbf{Failure Mode} & \textbf{Local Effect} & \textbf{Potential Severity} & \textbf{Recommendation} \\ 
\hline 
Traffic Detection & Sensor Failure & No traffic detected & Critical (10) & Implement dual-sensing technology with automatic failover. Ensure periodic testing of failover capability. \\ 
\hline 
Traffic Detection & Software Error & Incorrect traffic data processing & High (9) & Conduct extensive integration testing and add error-checking algorithms to validate data before processing. \\  
\hline 
Traffic Detection & Communication Failure & Delayed or no data transmission & High (8) & Implement redundant communication channels and improve error-handling procedures to detect and reroute data transmission automatically. \\  
\hline 
Traffic Detection & Configuration Error & Incorrect detection parameters set & Medium (7) & Standardize configuration procedures and introduce periodic configuration audits and validation tests. \\
\hline 
Traffic Detection & Physical Obstruction & Sensor input blocked or degraded & Medium (6) & Schedule regular maintenance checks and install environmental monitoring systems to alert staff to obstructions. \\ 
\hline 
Traffic Tracking & Loss of tracking data & Inability to continue monitoring detected traffic & High (8) & Implement data recovery protocols and redundant tracking mechanisms with automatic switching to backup systems. \\  
\hline 
Traffic Tracking & Erroneous velocity vector computation & Incorrect assessment of traffic movement & Medium (7) & Enhance algorithm accuracy with machine learning techniques and introduce additional validation checks at each computation stage. \\  
\hline 
Traffic Tracking & Sensor misalignment & Degraded tracking accuracy & Medium (6) & Implement regular calibration protocols and real-time monitoring of sensor alignment with alerts for deviations. \\ 
\hline 
Traffic Tracking & Communication delay & Late update of traffic data, potential for outdated tracking info & High (8) & Upgrade to faster data transfer technology and establish a secondary communication protocol for redundancy. \\  
\hline 
Traffic Tracking & Software crash & System stops tracking traffic altogether & High (9) & Develop a robust error handling framework and system recovery processes including auto-restart features. \\  
\hline 
Collision Evaluation & Incorrect threat assessment & Incorrect collision potential evaluation & High (8) & Integrate comprehensive machine learning algorithms to refine threat assessment based on real-world data continuously. \\  
\hline 
Collision Evaluation & Delayed data processing & Late collision threat response & High (8) & Optimize processing algorithms for efficiency and upgrade hardware for faster data handling capabilities. \\  
\hline 
Collision Evaluation & Sensor data corruption & Inaccurate threat data used for evaluations & High (8) & Implement stringent data integrity checks and use error-correcting codes for all incoming data streams. \\  
\hline 
Collision Evaluation & Software malfunction & System fails to evaluate collision threats & Critical (10) & Establish a routine for periodic software audits and develop robust error handling and recovery procedures. \\  
\hline 
Collision Evaluation & Communication failure with tracking component & No updated data received for evaluation & High (8) & Ensure reliable communication protocols are in place and implement fallback mechanisms for immediate switch-over when failures are detected. \\  
\hline 
Threat Prioritization & Incorrect ranking logic & Misordered threat prioritization & High (8) & Audit and optimize ranking algorithms annually and incorporate adaptive machine learning to enhance decision-making. \\  
\hline 
Threat Prioritization & Delay in processing updates & Outdated threat prioritization & High (8) & Invest in faster processing hardware and optimize real-time data handling capabilities to process updates instantaneously. \\  
\hline 
Threat Prioritization & Data synchronization errors & Inconsistent threat data used & High (8) & Implement a robust protocol for data verification and synchronization across all system components. \\  
\hline 
Threat Prioritization & Software glitches & Interruption in threat prioritization process & Critical (10) & Conduct frequent software testing and ensure updates are deployed; establish fail-safe modes for glitch detection. \\  
\hline 
Threat Prioritization & Communication breakdown with collision evaluation system & Lack of input for prioritization decisions & High (8) & Design a multi-channel communication strategy with automatic failover to backup channels in case of failure. \\  
\hline Maneuver Determination & Inaccurate threat assessment & Incorrect maneuver determination & High (9) & Continuously update and refine threat assessment algorithms based on latest data and simulation outcomes. \\  
\hline Maneuver Determination & Delay in maneuver computation & Delayed response to collision threats & High (9) & Streamline computation algorithms and invest in advanced computing hardware to minimize response times. \\  
\hline Maneuver Determination & Software bugs or glitches & Incorrect or no maneuver generated & Critical (10) & Implement comprehensive testing protocols, regular software audits, and updates to detect and rectify bugs. \\  
\hline 
Maneuver Determination & Communication latency with sensors & Outdated data used for maneuver planning & High (8) & Enhance the real-time data transmission system and implement a checking mechanism to confirm data freshness. \\  
\hline 
Maneuver Determination & System overload & System unresponsive or slow to determine maneuvers & Critical (10) & Design the system for scalability, implement load balancing, and prioritize critical data processing. \\  
\hline 
Maneuver Command & Command signal failure & No maneuver initiated or incorrect maneuver initiated & Critical (10) & Implement comprehensive signal integrity checks and establish multiple fallback communication protocols. \\  
\hline 
Maneuver Command & Delay in command execution & Delayed response in executing maneuvers & High (9) & Optimize system response time with real-time processing enhancements and streamline command execution paths. \\  
\hline 
Maneuver Command & Incorrect command data & Inappropriate maneuver based on incorrect or corrupted data & Critical (10) & Strengthen data validation processes at each stage and employ redundant data verification systems. \\  
\hline 
Maneuver Command & Software malfunction & Incorrect maneuver execution, potential system crash & Critical (10) & Establish a protocol for regular software testing, updates, and robust error handling systems. \\  
\hline Maneuver Command & Hardware failure & Inability to execute commands due to hardware malfunction & High (9) & Utilize high-reliability hardware components, conduct regular maintenance, and implement hardware health monitoring. \\  
\hline 
\end{longtable}

\end{document}